\documentclass[prx,twocolumn,superscriptaddress,longbibliography]{revtex4-2}

\usepackage{upgreek}                
\usepackage{xfrac}                  
\usepackage{graphicx}               
\usepackage{amsmath}                

\usepackage{xcolor}
\definecolor{customlinkcolor}{RGB}{46,48,146}

\definecolor{my_yellow}{RGB}{255,180,0}
\definecolor{my_dark_yellow}{RGB}{191,135,0}
\definecolor{my_green}{RGB}{130,200,0}

\usepackage[colorlinks=true,allcolors=customlinkcolor]{hyperref}    

\DeclareRobustCommand{\orcidicon}{
    \hspace{-2.8mm}                             
    \includegraphics[width=2.5mm]{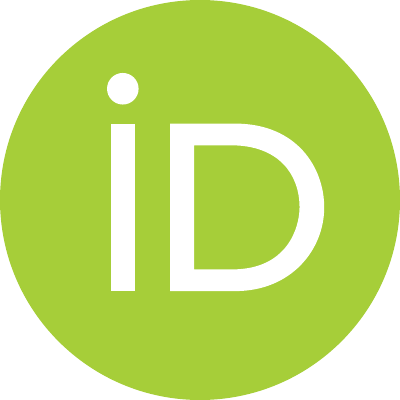} 
    \hspace{-1.3mm}}                            
\newcommand{\orcid}[1]{\href{https://orcid.org/#1}{\orcidicon}}
    \usepackage{cleveref}
    \crefname{equation}{Eq.}{Eqs.}
    \crefname{figure}{Fig.}{Figs.}
    \crefname{table}{Tab.}{Tabs.}
    \crefname{section}{section}{sections}

\begin{document}

\title{High-Field Optical Cesium Magnetometer for Magnetic Resonance Imaging}

\author{Hans St{\ae}rkind\orcid{0000-0002-6844-3305}}
\email{hans.staerkind@nbi.ku.dk}
\affiliation{Niels Bohr Institute, University of Copenhagen, Blegdamsvej 17, 2100 Copenhagen, Denmark}
\affiliation{Danish Research Centre for Magnetic Resonance, Centre for Functional and Diagnostic Imaging and Research, Copenhagen University Hospital - Amager and Hvidovre, Ketteg{\aa}rd All{\'e} 30, 2650 Hvidovre, Denmark}
\author{Kasper Jensen\orcid{0000-0002-8417-4328}}
\affiliation{Niels Bohr Institute, University of Copenhagen, Blegdamsvej 17, 2100 Copenhagen, Denmark}
\affiliation{School of Physics and Astronomy, University of Nottingham, University Park, Nottingham NG7 2RD, England, United Kingdom}
\author{J{\"o}rg H. M{\"u}ller\orcid{0000-0001-6984-0487}}
\affiliation{Niels Bohr Institute, University of Copenhagen, Blegdamsvej 17, 2100 Copenhagen, Denmark}
\author{Vincent~O.~Boer\orcid{0000-0001-6026-3134}}
\affiliation{Danish Research Centre for Magnetic Resonance, Centre for Functional and Diagnostic Imaging and Research, Copenhagen University Hospital - Amager and Hvidovre, Ketteg{\aa}rd All{\'e} 30, 2650 Hvidovre, Denmark}
\author{Eugene S. Polzik\orcid{0000-0001-9859-6591}}
\affiliation{Niels Bohr Institute, University of Copenhagen, Blegdamsvej 17, 2100 Copenhagen, Denmark}
\author{Esben T. Petersen\orcid{0000-0001-7529-3432}}
\affiliation{Danish Research Centre for Magnetic Resonance, Centre for Functional and Diagnostic Imaging and Research, Copenhagen University Hospital - Amager and Hvidovre, Ketteg{\aa}rd All{\'e} 30, 2650 Hvidovre, Denmark}
\affiliation{Section for Magnetic Resonance, DTU-Health Tech, Technical University of Denmark, Oersteds Plads, Building 349, 1st floor, 2800 Kgs Lyngby, Denmark}

\date{\today}

\begin{abstract}
We present a novel high-field optical quantum magnetometer based on saturated absorption spectroscopy on the extreme angular-momentum states of the cesium D\textsubscript{2} line. With key features including continuous readout, high sampling rate, and sensitivity and accuracy in the ppm-range, it represents a competitive alternative to conventional techniques for measuring magnetic fields of several teslas. The prototype has four small separate field probes, and all support electronics and optics are fitted into a single 19-inch rack to make it compact, mobile, and robust. The field probes are fiber coupled and made from non-metallic components, allowing them to be easily and safely positioned inside a 7~T MRI scanner. We demonstrate the capabilities of this magnetometer by measuring two different MRI sequences, and we show how it can be used to reveal imperfections in the gradient coil system, to highlight the potential applications in medical MRI. We propose the term EXAAQ (EXtreme Angular-momentum Absorption-spectroscopy Quantum) magnetometry, for this novel method.
\end{abstract}

\maketitle

\section{Introduction}
High magnetic fields play a critical role in many areas of science and technology. These include fundamental physics \cite{Battesti2018}, materials science \cite{Roy2017,Caridad2018}, mass spectrometry \cite{Marshall1998,Gorshkov1998}, particle accelerators \cite{Bottura2012,Bottura2016}, nuclear fusion \cite{Savary2010,Zhai2021}, magnetic levitation \cite{Berry1997,Wang2002,Floegel-Delor2019}, chemistry \cite{Bothwell2011,Wikus2022} and medical imaging \cite{Harisinghani2019,Lvovsky2013}.

When measuring high magnetic fields, of more than \mbox{$1\:\mathrm{T}$}, four different conventional techniques can be employed, each with their own advantages and disadvantages \cite{Keller_Metrolab,Bottura2009,Lenz2006}. \textit{Nuclear magnetic resonance (NMR) magnetometers}, using protons, deuterons, helions, or heavier nuclei, measure the scalar magnetic field magnitude and are superior in terms of accuracy and sensitivity, but typically function in a pulsed way and require high field homogeneity \cite{DeZanche2008,PT2026_Metrolab,Nikiel2014,Barmet2010}. \textit{Hall probes} provide continuous measurements of a vector component of the magnetic field, with high spatial resolution, but with lower sensitivity \cite{3MH6-E_Senis}. \textit{Fluxmeters} are very sensitive, but they are relative devices that can only measure changes in magnetic fields along a specific direction and will drift when operated continuously \cite{Pedersen2009,Adaikkan2022,FDI2056_Metrolab}. \textit{Magneto-optical Faraday rotation magnetometers} provide fast optical measurements of a vector component of the magnetic field, but with poor sensitivity \cite{Nakamura2013,Tsuli-lio2001}.

Optical magnetometry based on measuring the Zeeman shift of alkali D lines provides a fifth approach \cite{George2017,Ciampini2017,Keaveney2019,Klinger2020}, but the real potential of this method has still not been demonstrated, and no practical applications have been explored yet. In this work we build upon the recent fundamental advances described in \cite{Staerkind2023}, to realize a novel prototype optical magnetometer that gives a fast and continuous measurement of the scalar magnetic field magnitude, with a high accuracy and sensitivity. The optical probes provide easy and interference-free operation, with minimal electromagnetic disturbances in the measured volume. Overall, this provides a compelling new technique for measuring high magnetic fields that may prove advantageous compared to conventional methods for certain applications, and even enable new ones. Like NMR magnetometers, this optical magnetometer is a quantum sensor that achieves absolute accuracy by exploiting the fact that all atoms of a given species are perfectly identical. This gives a high robustness against component ageing, and ambient fluctuations.

The method is based on fast tracking of a magnetic-field-dependent near-infrared transition in cesium, combining sideband spectroscopy, saturated absorption spectroscopy, and frequency modulation (FM) spectroscopy. A similar approach has originally been proposed in \cite{Brauer2012}. Here we present the first ever physical realization of such a device.

\begin{figure*}[t]
    \centering
    \includegraphics{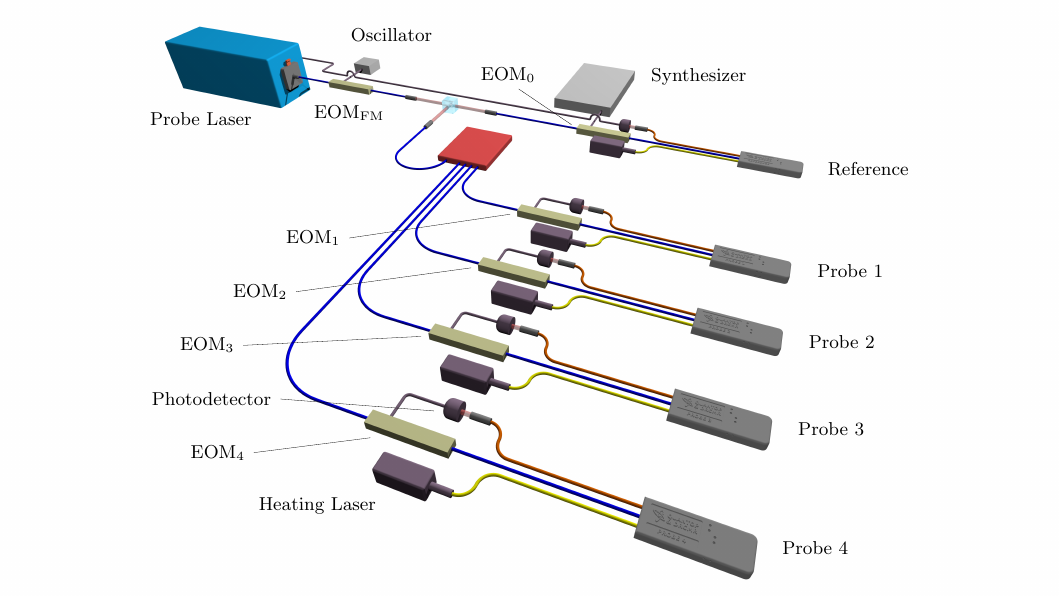}
    \caption{Optical setup. Probe laser light is modulated by EOM\textsubscript{FM}, split by a free-space beam splitter, and then further split by a 1x4 fiber splitter, for a total of five different paths. Each path is again modulated by an EOM, before the probe light is sent to the reference probe at 0 T and probes 1--4 in the MRI scanner. Colored cables represent optical fibers, and black cables represent the analog feedbacks realizing the accurate measurement scheme.}
    \label{fig:Setup}
\end{figure*}

We have developed the prototype specifically to monitor the magnetic field inside a magnetic resonance imaging (MRI) scanner. As described in \cite{Barmet2008}, accurate magnetic field data can be used in the MRI image reconstruction to improve image quality, with potential impact in both medical research and clinical diagnostics. To facilitate the development and testing of the prototype in a hospital setting, it has been made to fit into a 19-inch rack, making it compact, mobile, robust, and easy to quickly turn on and operate. This has been necessary since the extreme magnetic-field conditions inside an MRI scanner are not easily reproduced in a typical lab environment. The prototype is characterized and tested in a Philips Achieva \mbox{$7\:\mathrm{T}$} MRI scanner. We benchmark the performance in terms of sensitivity and bandwidth, and show measurements of an echo-planar imaging (EPI) sequence and a spiral imaging sequence. These measurements lead to an investigation of clearly detectable nonlinearities and instabilities in the gradient coil system, indicating that the prototype already performs at a level where it can serve as a useful tool for investigating the performance of the MRI scanner. Finally we discuss the absolute accuracy of the method.

With this EXAAQ (EXtreme Angular-momentum Absorption-spectroscopy Quantum) device being the very first of its kind, there are numerous opportunities to further improve its reliability, sensitivity, and accuracy. While we will continue to develop and improve the prototype towards applications in MRI, we note that a modified design may find applications in other fields. This could include: magnetic diagnostics in steady state fusion experiments, where the drift of fluxmeters or radiation damage to Hall probes becomes a problem \cite{Strait2008,Quercia2022}; accurate control of dipole magnets in particle accelerators, through continuous optical monitoring as originally proposed in \cite{Brauer2012}; or quench detection in (high-temperature) superconducting magnets as an alternative to ``quench antennas'' or Hall probe arrays \cite{Marchevsky2021}.

\section{Method}
\subsection{The optical transition}
The magnetometer works by continuously tracking the frequency shift of the cesium-133 D\textsubscript{2} line. Specifically it measures the shift $\Delta\nu_+$, of the $\sigma_+$ transition between the extreme angular-momentum states, shown in \cite{Staerkind2023} to have a magnetic field dependence of
\begin{align}
    \Delta\nu_+ = \gamma_1 B + \gamma_2 B^2.
    \label{eq:Freq_shift}
\end{align}
Here $\gamma_1=13.994\,301(11)\:\mathrm{GHz/T}$ is the linear magnetic frequency shift of the transition, \mbox{$\gamma_2=0.4644(35)\:\mathrm{MHz/T^2}$} is the quadratic diamagnetic shift of the transition, and $B$ is the magnitude of the magnetic field. This expression is modified to
\begin{align}
    \Delta\nu_+ = \gamma_0 + \gamma_1\zeta B + \gamma_2\zeta^2 B^2,
    \label{eq:Freq_shift_in_practise}
\end{align}
such that $B$ is defined as in the absence of the probe, by taking into account the probe induced field shift \mbox{$\zeta = 1 + 0.49(50)\times 10^{-6}$}. A small frequency measurement offset $\gamma_0$, depending on the measurement method, is also included in \cref{eq:Freq_shift_in_practise}.

With the linear magnetic frequency shift of about 14~GHz/T dominating \cref{eq:Freq_shift_in_practise}, we have a frequency shift of about 98 GHz in the 7 T MRI scanner, where we test the magnetometer.

\subsection{Setup}
\begin{figure}[htb]
    \centering
    {\includegraphics[width=\linewidth]{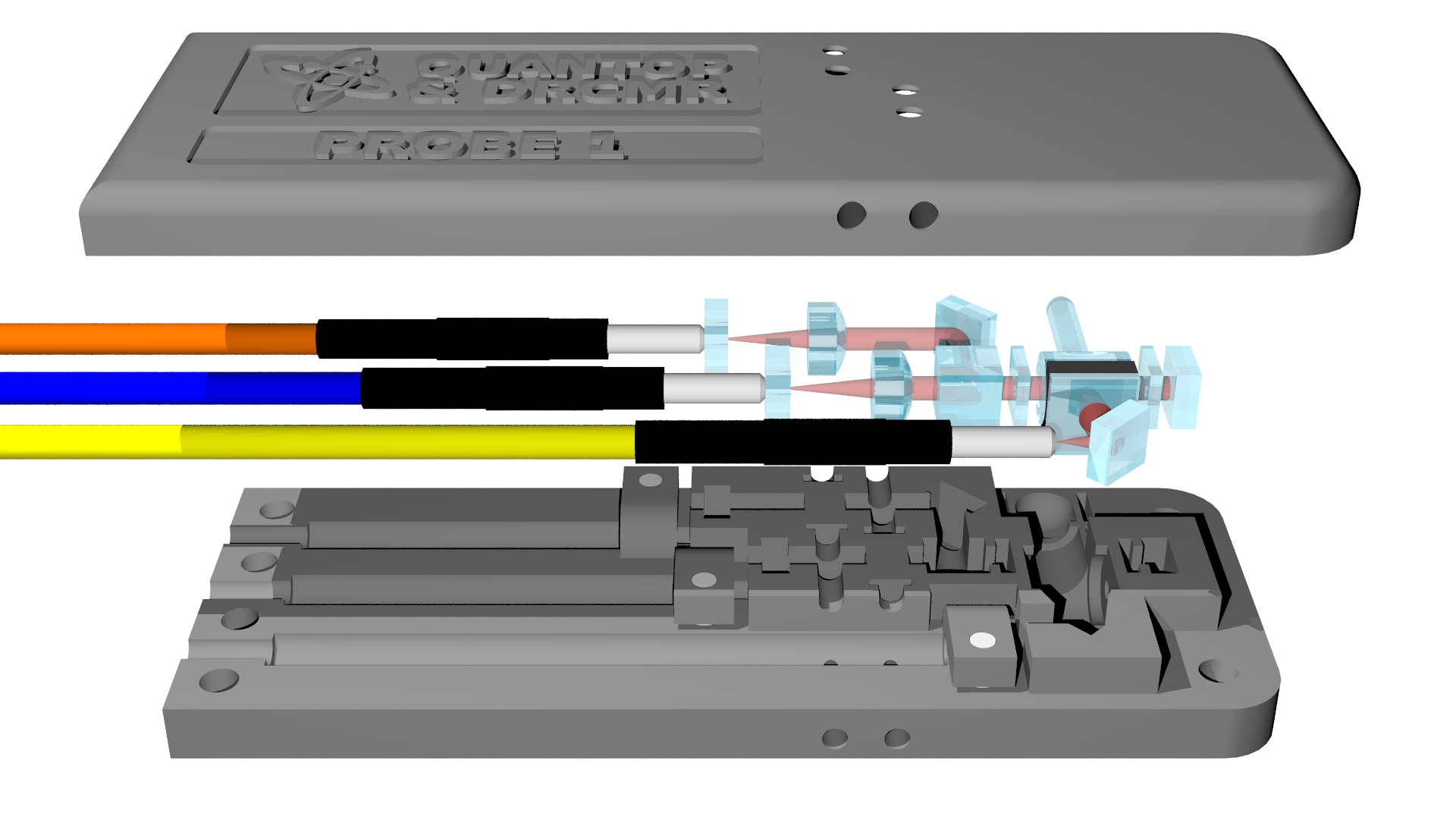}}
    \caption{Exploded view of the optical probe assembly. Probe light comes from the blue fiber, and exits through the orange fiber. High-power laser light from the yellow fiber heats the cesium vapor cell, to increase the cesium atomic density. The assembled probes dimensions are $90\times 33\times 10\: \mathrm{mm}^3$.}
    \label{fig:Exploded_assembly}
\end{figure}

A schematic view of the optical setup is shown in \cref{fig:Setup}. The probe laser is a Toptica DL Pro, 852 nm, external cavity diode laser (ECDL). The probe laser light is first modulated at $\nu_\mathrm{FM} =  5.32\:\mathrm{MHz}$ by an electro-optical modulator (EOM), which we label EOM\textsubscript{FM}. This modulation is necessary for performing FM spectroscopy as described in \cref{sec:Feedback}. The laser beam is then split in two paths, and one path is further split into four, for a total of five different paths. In each path the probe light is again modulated by an EOM to generate strong sidebands for sideband spectroscopy as detailed in \cref{sec:Frequency_shift_measurement}. In each path the light is then sent to a 3D-printed probe, containing all the optics for performing saturated absorption spectroscopy \cite{Schmidt1994} on a laser-heated spectroscopy cell. An exploded view of the probe assembly is shown in \cref{fig:Exploded_assembly}.
Apart from the input fiber for the probe light, the probes also have an input fiber for the heating laser beam, and an output fiber for returning the probe light to a photodetector. The fibers are 19~m long allowing the rack to be placed at a safe distance from the measurement points.

The probes are labeled ``probes 1--4'', and ``reference'' as seen in \cref{fig:Setup}. Probes 1--4 are used to measure at four different positions inside the MRI scanner, and the reference probe is placed in a magnetic shield in the rack. Probes 1--4 are all configured with $\sigma_+$ polarized probing light, whereas the reference uses linearly polarized light to prevent the transition frequency from shifting in small residual field fluctuations inside the magnetic shield. For details on the optical probe design see \cite{Staerkind2023}.

Each of probes 1--4 are heated with 1.2~W of optical power to a temperature of about 44~\textdegree C, and uses about 180~\textmu W of probing light.
The reference is heated with 250~mW to a temperature of about 35~\textdegree C, and uses about 220~\textmu W of probing light.

\subsection{Frequency shift measurement}
\label{sec:Frequency_shift_measurement}
With the reference in the magnetic shield at 0~T, and the four probes in the MRI scanner at 7~T, we initialize the magnetometer using a two-step procedure.

In the first step the probe laser frequency is stabilized 97.5~GHz above the 0~T transition. This is achieved by driving the modulator in the reference path, EOM\textsubscript{0}, with a frequency $\nu_0 = 19.5\:\mathrm{GHz}$ and power optimized for the fifth optical sidebands. The laser frequency is adjusted such that the lower fifth sideband is resonant with the 0 T transition, and the laser is then stabilized at this frequency by an electronic feedback to the laser current controller. Feedback methods are described in \cref{sec:Feedback}.

\begin{figure}[htb]
\centering
\includegraphics{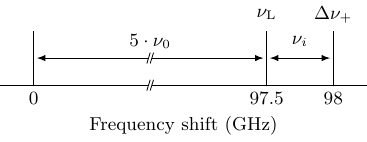}
\caption{The sideband spectroscopy frequency shift measurement scheme illustrated. The probe laser frequency $\nu_\mathrm{L}$ is stabilized 97.5~GHz above the 0~T transition by spectroscopy with the lower fifth sideband of $\nu_0=19.5\:\mathrm{GHz}$ using the reference probe. For each of the probes $i=\{1,2,3,4\}$, a first upper sideband of about 0.5~GHz follows the 7~T transition. The total frequency shift, $\Delta\nu_+$, is then found as  $5\cdot\nu_0 + \nu_i$.}
\label{fig:Frequency_shift_measurement}
\end{figure}

In the second step the modulator EOM\textsubscript{$i$}, in each of the four probe paths $i=\{1,2,3,4\}$, is used to modulate the probe light with a frequency of about $\nu_i=0.5\:\mathrm{GHz}$ and power optimized for the first sidebands. The modulation frequency is fine-tuned such that the first upper sideband is resonant with the 7~T transition, and it is then stabilized using an electronic feedback to the voltage controlled oscillator (VCO) driving EOM\textsubscript{$i$}. Now the total frequency shift from \cref{eq:Freq_shift_in_practise} can be found as
\begin{align}
    \Delta\nu_+ = 5\cdot\nu_0 + \nu_i,
    \label{eq:Calc_freq_shift}
\end{align}
as shown in \cref{fig:Frequency_shift_measurement}.
A measurement of $\nu_i$ then corresponds directly to a magnetic field measurement through \cref{eq:Calc_freq_shift,eq:Freq_shift_in_practise}.

The feedback will, once initialized, make sure that the modulation frequency $\nu_i$ follows changes in the resonance frequency, corresponding to changes in the magnetic field, such that \cref{eq:Calc_freq_shift} remains valid at all times, except for a small possible error due to the finite speed of the feedback. Each of the four probes in the MRI scanner has its own feedback, so that an independent measurement is made by each probe.

The VCOs used to drive EOM\textsubscript{1--4} can produce frequencies $\nu_i$, in the range from 320 to 760 MHz, giving the magnetometer a dynamic range of about $\pm$ 15 mT. With the probes at a distance of 15 cm from the MRI magnet isocenter, this allows for measurements of magnetic field gradients up to 100 mT/m, which is more than most MRI gradient coil systems can produce. It should be noted that the system can easily be reconfigured to measure around a completely different field strength, e.g.\ for 3 T or 1.5 T MRI scanners, by simply changing $\nu_0$.

\subsection{Feedback}
\label{sec:Feedback}
A saturated absorption spectroscopy signal is symmetric around the resonance frequency, and as such is not directly useful in a feedback. To make a feedback, an asymmetric error signal must be generated from the probe laser light, such that positive and negative frequency detunings are easily distinguished. This is achieved using FM spectroscopy~\cite{Bjorklund1983}. In FM spectroscopy the probing light is modulated at a frequency similar to the linewidth of the transition. The sidebands and carrier pick up different phase shifts, and interfere to give a light intensity that oscillates at the modulation frequency with phase and amplitude depending on the sign and magnitude of the detuning. The probe laser light is modulated at $\nu_\mathrm{FM} =  5.32\:\mathrm{MHz}$, using EOM\textsubscript{FM}, since this frequency is similar to the transition linewidth of 5.2~MHz \cite{Young1994,Steck2023}, and also matches the free spectral range of the fiber etalons, such that their contribution to the error signal is suppressed.

\begin{figure}[htb]
    \centering
    \scriptsize
    \includegraphics{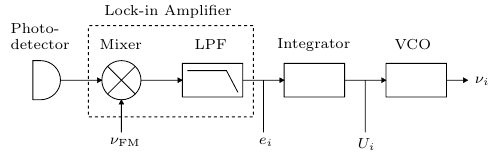}
    \caption{The analog feedback used to control the frequencies of EOM\textsubscript{1--4}. The signal from the frequency-selective photodetector is passed through a lock-in amplifier, to generate the error signal $e_i$. This is integrated by an analog op-amp integrator, to produce a voltage $U_i$, controlling a VCO, producing a frequency $\nu_i$, which is sent to EOM\textsubscript{$i$}.}
    \label{fig:Feedback}
\end{figure}

Each photodetector is designed as a parallel RLC circuit with a resonance of $\nu_\mathrm{FM}$ and a Q-factor of about 15, to reduce sensitivity outside the relevant narrow frequency band. The signal from the photodetector is sent to a lock-in amplifier, consisting of an analog mixer and a 1.9 MHz low pass filter (LPF), to produce the error signal $e_i$. The error signal is a voltage which is positive for negative detunings, and negative for positive detunings. The error signal is sent to an integrator, which controls the VCO with a voltage $U_i$, to produce a frequency $\nu_i$, which is sent to EOM\textsubscript{$i$}. This is all shown in \cref{fig:Feedback}.

\begin{figure}[tb]
    \centering
    \includegraphics{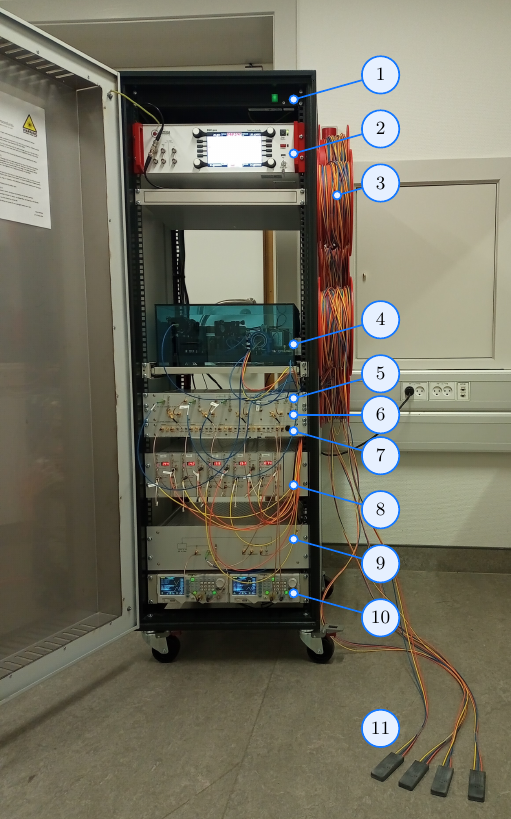}
    \caption{EXAAQ magnetometer prototype. 1: On/off switch. 2: Toptica DLC Pro laser controller. 3: Cable reels with 19 m of fiber. 4: Optical breadboard with free-space beam splitting, synthesizer and amplifier for $\nu_0$, EOM\textsubscript{0}, and magnetic shield with reference. 5: 1x4 fiber splitter, amplifiers for $\nu$\textsubscript{1--4}, and EOM\textsubscript{1--4}. 6: VCOs. 7: Lock-in amplifiers and integrators for probes 1--4. 8: Subrack with photodetectors and heating lasers. 9: Probe laser, EOM\textsubscript{FM}, and lock-in amplifier and integrator for reference. 10: Oscillators for $\nu_\mathrm{FM}$ and VCO voltage scanning. 11: Probes 1--4.}
    \label{fig:Photo_markings}
\end{figure}

\begin{figure}[thb]
  \centering
  \includegraphics{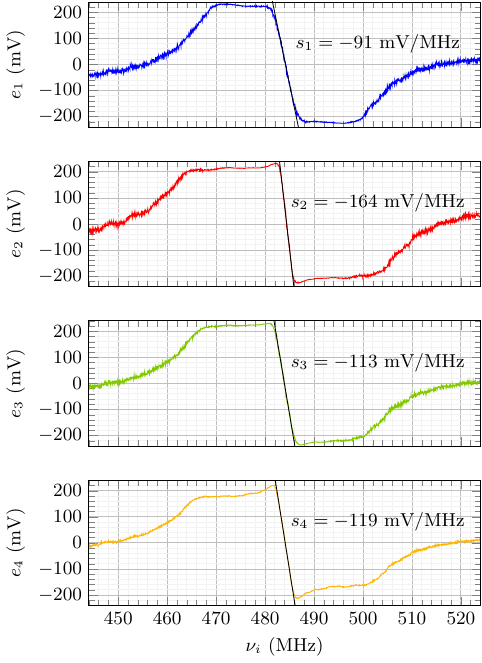}
  \caption{Error signals $e_i$, recorded as a function of $\nu_i$. This range of frequencies is produced by sweeping $U_i$ about 2~V. A fit to the linear region around resonance, $e_i=0$, returns the slopes $s_i$.}
  \label{fig:02_error_calibration}
\end{figure}

As an example, we examine what happens when the magnetic field suddenly increases at the position of probe~$i$, resulting in a negative detuning of the first upper EOM\textsubscript{$i$} generated sideband. This will give a positive $e_i$, which will be integrated to give a rising $U_i$, increasing $\nu_i$ until the negative detuning is fully compensated, and $e_i$ is again zero. This corresponds to the first upper EOM\textsubscript{$i$} generated sideband again being on resonance with the atomic transition, and \cref{eq:Calc_freq_shift} again being valid.

The laser frequency stabilization using the reference probe at 0 T and a fixed $\nu_0$ is similar, except that there is no VCO. In this case the integrator directly controls the laser current, compensating, e.g., mechanical disturbances and temperature fluctuations, to keep the laser frequency constant at all times.

\cref{eq:Calc_freq_shift_general} can be generalized to situations where the field is rapidly changing, and the first upper sideband is not exactly on resonance. In this case we have
\begin{align}
    \Delta\nu_+ = 5\cdot\nu_0 + \nu_i - e_i/s_i,
    \label{eq:Calc_freq_shift_general}
\end{align}
where $s_i$ is the slope of the error signal as a function of detuning.

For the measurements presented in the following, $\nu_i$ is found from $U_i$, using a voltage-to-frequency mapping of the VCOs. The VCOs behave as low pass filters with a time constant of about 4~\textmu s, corresponding to a bandwidth of about 40~kHz. These numbers also include response times of photodetectors and lock-in amplifiers.

\begin{figure}[htb]
  \centering
  \includegraphics{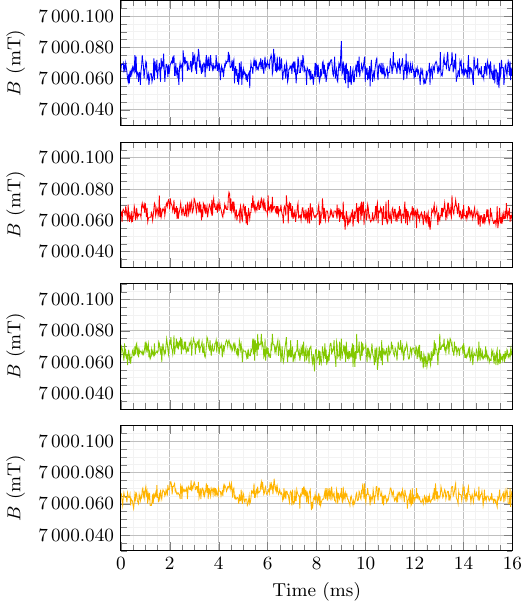}
  \caption{Example measurements, in the magnetic field of the MRI scanner. The sampling rate is 40~kHz. The field is assumed to be a perfect $7\,000.066$~mT. Colors blue, red, green, and yellow corresponds to probes 1--4, respectively. Probe 4 is the one that has the least noise, with a standard deviation of 3.9~\textmu T, over a 1 second measurement. Noise common among the probes is attributed to imperfect frequency stabilization of the probe laser.}
  \label{fig:03_Silent_time}
\end{figure}

\begin{figure}[htb]
  \centering
  \includegraphics{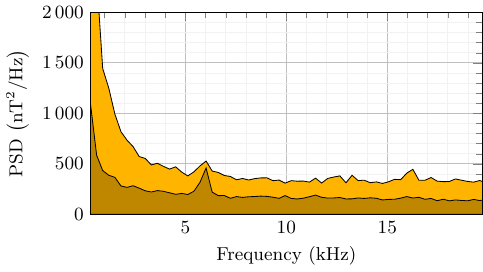}
  \caption{The PSD of a measurement with probe 4. The shaded lower area shows the noise of the data acquisition system.}
  \label{fig:accum_PSD}
\end{figure}

\begin{figure*}[htb]
  \centering
  \includegraphics{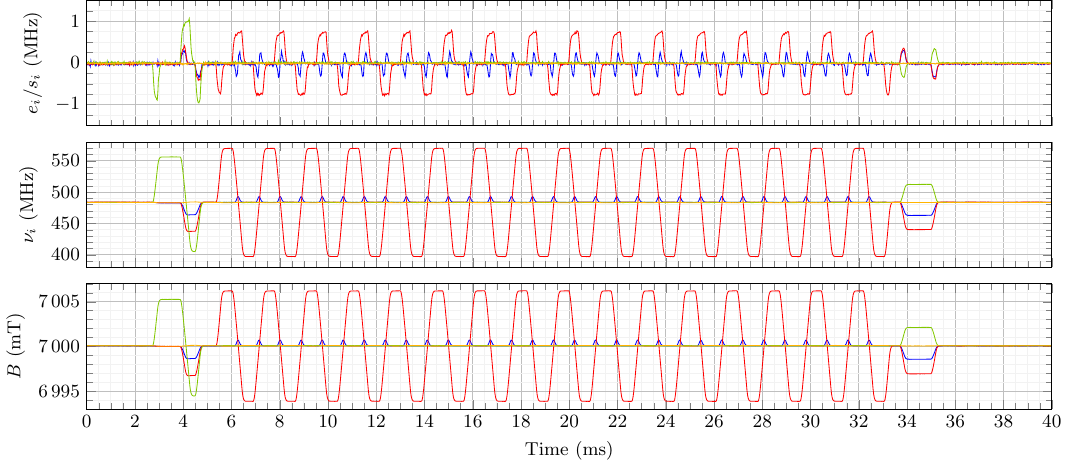}
  \caption{Magnetic field measurements during an EPI sequence with probes 1--4, in colors blue, red, green, and yellow, respectively. In the upper plot are shown the differences between the first upper sidebands and the atomic resonances. In the middle plot are shown the VCO frequencies. In the lower plot are shown the magnetic fields calculated using \cref{eq:Freq_shift_in_practise,eq:Calc_freq_shift_general}. The reader familiar with MRI sequence development will recognize slice selection, prewinder, k-space coverage, and spoiler \cite{Bernstein2004}.}
  \label{fig:05_EPI_recording}
\end{figure*}

\section{Results}
\subsection{19-inch rack integration}
To make a compact device that can be operated in a hospital setting, the setup is divided in several modules and mounted in a 19-inch rack. This also makes it easy to move around and close off, when not in use. To make it robust, all optical parts of the setup are directly connected by fibers; except for the beam splitting after EOM\textsubscript{FM}, which is realized with free-space optics, to allow for optical power adjustments. The device is shown in \cref{fig:Photo_markings}.

During measurements, the device is placed in the control room of the MRI scanner, the probes enter the radio frequency (rf) shielded scanner room through a wave\-guide, and are placed in the bore of the scanner in a 3D printed plastic grid, with a precision of about 1 mm.

\subsection{Calibration and sensitivity characterization}
With the laser frequency stabilized, and the probes in the MRI scanner bore, the error signal slopes $s_i$, from \cref{eq:Calc_freq_shift_general}, are determined. This is done by sweeping the VCO control voltage $U_i$, and thereby the frequency $\nu_i$, while recording the error signal $e_i$. A straight line is fitted to the linear region around resonance, as shown in \cref{fig:02_error_calibration}, to find the slopes $s_i$. 

Once the probe feedbacks are initialized, the VCO frequencies $\nu_i$ are continuously regulated so that \cref{eq:Calc_freq_shift_general} remains valid as long as the error signals $e_i$ stay within the approximately linear region of about $\pm 1.5$ MHz.

To calibrate the probes in absolute terms, we first measure the magnetic field in the MRI scanner by NMR magnetometry. Placing a spherical sample of ultrapure water in the scanner and measuring a free induction decay following nuclear excitation, we find the magnetic field in the center of the scanner to be $7.000\,066$ T \cite{Tiesinga2021,Staerkind2023}. We then measure $\Delta\nu_+$ as defined in \cref{eq:Calc_freq_shift_general}, with each of the four probes for 1 second. We assume that the high magnetic field generated by the superconducting MRI coil is completely unchanged during all measurements.  Using \cref{eq:Freq_shift_in_practise} we then calculate a $\gamma_0$ for probes 1--4 of $0.07$, $0.19$, $-0.26$, and $-0.19$~MHz, respectively. Using these values we measure example magnetic field traces as shown in \cref{fig:03_Silent_time}. Here a sampling rate of 40 kHz is used, corresponding to a highest detectable (Nyquist) frequency of 20 kHz. Noise common among the probes is likely due to imperfections in the frequency stabilization of the probe laser.

Over a 1 second measurement we measure traces with standard deviations of 5.0, 4.3, 4.3, and 3.9~\textmu T, respectively. This constitutes a sub-ppm resolution of the magnetometer. We calculate the power spectral density (PSD) as shown in \cref{fig:accum_PSD} for the measurement with probe 4, which is the one that performs the best, and note that the noise is worst below 3~kHz. The sensitivity can be calculated as the square root of the PSD, and has a root mean square value of 28 nT/$\sqrt{\mathrm{Hz}}$ in the 0--20~kHz band. It is worth noting that the noise of the data acquisition system used for measuring $U_i$, also shown in \cref{fig:accum_PSD}, corresponds to a standard deviation of 2.5~\textmu T, so this is currently a limiting factor for the sensitivity. Over a 20~minutes period we observe drifts of 23, 17, 34, and 24~\textmu T, respectively, i.e.\ a long term stability of a few ppm.

\begin{figure*}[htb]
  \centering
  \includegraphics{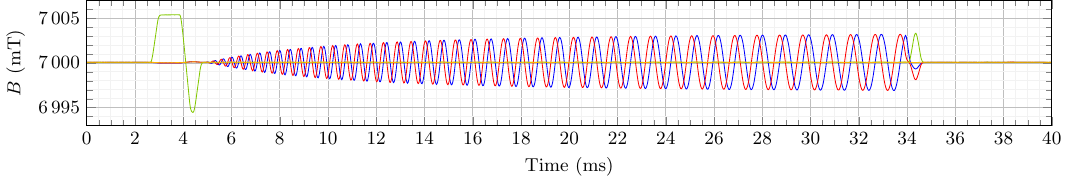}
  \caption{Magnetic field measurements during a spiral imaging sequence with probes 1--4, in colors blue, red, green, and yellow, respectively.}
  \label{fig:08_Spiral_recording}
\end{figure*}

\subsection{MRI sequence measurements}
We demonstrate the functioning of the prototype by measuring the field in four different positions inside the MRI scanner while scanning. During an MRI scan three orthogonal gradient coils are playing a sequence of carefully controlled magnetic field gradients. Here we investigate typical 2D MRI sequences: First a slice selecting gradient in one direction is played during nuclear rf excitation. This is followed by a series of gradients in the orthogonal plane, producing changing spatial nuclear spin phase rolls in the excited slice. These changing phase rolls determine the trajectory though k-space during inductive readout of the nuclear rf signal. During the final image reconstruction the acquired k-space image is Fourier transformed to give the actual image \cite{Bernstein2004}.

We define the coordinate system $(x,y,z)$ of the MRI scanner with $x$-axis along the down/up direction, $y$-axis along the left/right direction, and $z$-axis along the field direction. The origin $(0,0,0)$ corresponds to the isocenter of the MRI scanner. The probes 1--4 are placed in positions $(15\:\mathrm{cm},0,0)$, $(0,15\:\mathrm{cm},0)$, $(0,0,15\:\mathrm{cm})$, and $(0,0,0)$, respectively. In this way probes 1 and 2 measure the $(x,y)$-gradients defining the k-space trajectory, probe 3 measure the slice selecting $z$-gradient and 4 should ideally measure a constant field.

To demonstrate that the technique works well, even for rapidly changing fields, an EPI sequence with maximum gradient strength (39.87 mT/m) and maximum slew rate (198.38 mT/m/ms) is played on the MRI scanner while measuring with the four probes. The measured $e_i/s_i$ and $\nu_i$, and the magnetic fields $B$, calculated using \cref{eq:Freq_shift_in_practise,eq:Calc_freq_shift_general}, are shown in \cref{fig:05_EPI_recording}. Importantly, it is seen here that the error signals stay within the approximately linear region, demonstrating that even for the fastest gradient switching, that the MRI scanner can produce, the method still works.

The trajectory though k-space is given by the integral of the gradients, so the EPI sequence covers k-space through a zigzag line-by-line trajectory. Different sequences use (among other things) different k-space trajectories to adjust e.g.\ tissue contrast, field-of-view, or scan duration. In \cref{fig:08_Spiral_recording} we show a recording of a spiral imaging sequence, which covers k-space in an outwards spiralling trajectory, by playing out-of-phase oscillating $x$- and $y$-gradients with increasing amplitude.

These two examples clearly illustrate that the prototype works well and represents a valuable tool for the MRI scientist who needs a direct measurement of the magnetic field inside the scanner.

\subsection{Gradient imperfection detection}
Since the k-space trajectory is determined by the field gradients, inaccurate knowledge of the gradients will result in an inaccurate k-space image. Errors in the k-space image will, in turn, translate into artifacts or blurring in the Fourier transformed actual anatomical image. For this reason accurate measurements of the magnetic field gradients during an MRI sequence, can be used for k-space trajectory corrections, which ultimately can lead to improved MRI image quality.

Many different factors may contribute to deviations from the desired magnetic field during a sequence: Crosstalk among the coils and other nearby conducting material --- i.e.\ eddy currents, imperfect design of the coil system and the associated current controllers, heating of the coils during extended operation, etc.

Two observations in the obtained MRI sequence data warrant further investigation: For probe 2 in the EPI sequence measurement, we calculate a maximum gradient strength which is about 200~\textmu T higher than expected. For probe 2 in the spiral sequence measurement, decaying oscillations, with amplitudes up to about 10 \textmu T, are observed immediately after the sequence has finished, where the field should ideally be completely stable. These observations are not visible in the zoomed out views of \cref{fig:05_EPI_recording,fig:08_Spiral_recording}. To further investigate and clearly display these disagreements we reposition probes 1--4 to positions $(0,-15\:\mathrm{cm},0)$, $(0,-7.5\:\mathrm{cm},0)$, $(0,+7.5\:\mathrm{cm},0)$, and $(0,+15\:\mathrm{cm},0)$, respectively, and play the two sequences again.

\begin{figure}[htb]
  \centering
  \includegraphics{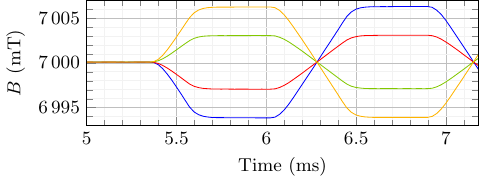}
  \caption{A small section of the EPI sequence shown in \cref{fig:05_EPI_recording}, recorded with the probes distributed along the $y$-axis.}
  \label{fig:07_EPI_recording}
\end{figure}

\begin{figure}[htb]
  \centering
  \includegraphics{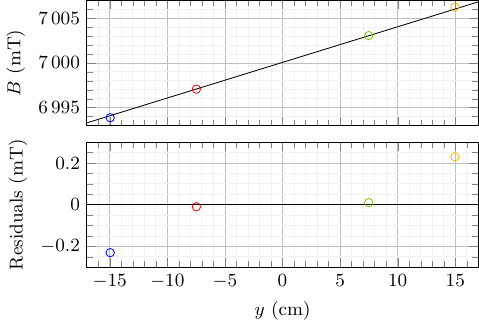}
  \caption{Nonlinearity in the magnetic field of the $y$-gradient coil. In the upper plot the magnetic field values of the first gradient plateau in \cref{fig:07_EPI_recording} are shown, along with an ideal 39.87~mT/m gradient. In the lower plot the residuals between the measured field and the ideal gradient are shown. At $\pm 15$~cm we see deviations of about 4 \%.} 
  \label{fig:07_EPI_recording_spatial}
\end{figure}

\begin{figure}[htb]
  \centering
  \includegraphics{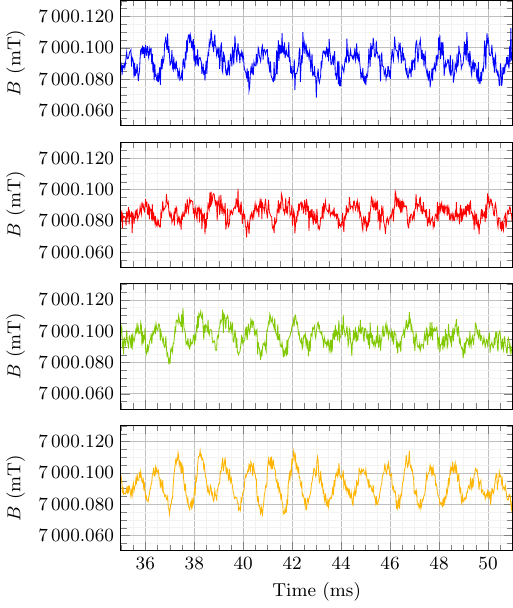}
  \caption{Measurement of the first 16 ms after the spiral sequence shown in \cref{fig:08_Spiral_recording}, recorded with the probes distributed along the y-axis. Blue, red, green, and yellow traces corresponds to positions $-15$, $-7.5$, $+7.5$, and $+15$ cm respectively. Oscillating eddy currents are clearly seen. Notice how the measurement has drifted about 20--30~\textmu T, since the calibration done 15 minutes earlier.}
  \label{fig:10_Spiral_recording}
\end{figure}
With the new probe positions we first have a closer look at the two first maximum gradient pulses in the EPI sequence in \cref{fig:07_EPI_recording}. In \cref{fig:07_EPI_recording_spatial} we plot the magnetic field values at the first gradient plateau for the four probes along with the field of an ideal 39.87~mT/m gradient. We also plot the residuals between the data and the idealization, and what is seen is a clear nonlinearity of the magnetic field generated by the $y$-gradient coil. To have nonlinearities like this is expected, and the order of magnitude --- about 4\% at 15 cm from the isocenter --- is comparable to what is e.g.\ found in \cite{Markl2003}.

\begin{figure}[thb]
  \centering
  \includegraphics{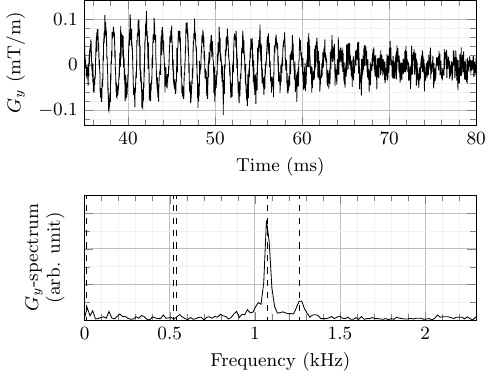}
  \caption{Magnetic field gradient along the $y$-axis calculated using the measurements by probes 1 and 4, shown partly in \cref{fig:10_Spiral_recording}. In the upper plot is shown the time domain signal starting directly after the spiral sequence has finished, and ending 45 ms later when the decaying oscillations are no longer visible. In the lower plot is shown the corresponding spectrum. The dashed lines shows the known oscillations that the ecc system tries to compensate.}
  \label{fig:10_Spiral_differential}
\end{figure}

For the spiral sequence we start by looking at the field during the first 16 ms after the sequence in \cref{fig:10_Spiral_recording}. Clear oscillations are seen. Highest amplitudes are seen for probes 1 and 4, which are the farthest away from the isocenter, and opposite amplitudes are seen for probes 1 and 2 compared to probes 3 and 4. This indicates an oscillating gradient along the $y$-axis. We calculate this magnetic field gradient using the measurements of probes~1 and 4, as shown in \cref{fig:10_Spiral_differential}. A spectral analysis of this signal, also displayed in \cref{fig:10_Spiral_differential}, shows how the signal clearly contains a strong component of about 1.071~kHz and a weaker component of about 1.263~kHz. These same two frequencies are found in the eddy current compensation (ecc) system of the MRI scanner. So what we see here are the remaining oscillations that the ecc system has not managed to fully compensate. Such oscillating instabilities have their origin in mechanical vibrations of the MRI scanner, excited by the strong forces from the switching gradient coils. In \cite{Winkler2018} they are described as ``vibrational eddy currents''.

\section{Discussion}
\subsection{Accuracy}
\label{sec:Discussion_Accuracy}
If we assume an error-free determination of $\gamma_0$ and $\nu_+$, then we can find $B$ from \cref{eq:Freq_shift_in_practise} with an accuracy of 13~\textmu T at 7~T. This is the scientific limit given by the best available data for the optical transition, and it is completely dominated by the uncertainty in $\gamma_2$.

A virtually error-free determination of $\gamma_0$ and $\nu_+$ is however not realized, as drifts up to \mbox{0.48 MHz} are observed, corresponding to about 10\% of the natural linewidth of the transition. This translates to a measurement drift of \mbox{34 \textmu T}, i.e.\ an accuracy about 5 ppm. The initially determined $\gamma_0$ for the four probes similarly have a spread of \mbox{0.45 MHz}, confirming that this is the level of accuracy achievable with the current hardware.

One possible explanation for such measurement drifts is temperature changes of the VCOs. The temperature dependence of the VCOs is reported to be up to \mbox{0.14 MHz/\textdegree C}, which means that several degrees of temperature changes would significantly change the measurement. After being switched on, the rack is allowed at least an hour to warm up, and settle on a temperature before measurements, for this reason.

A second explanation for drifts is associated with the laser-frequency stabilization. In the absorption spectrum, the saturation peak of the reference is located at a sloped background due to the other Doppler-broadened nearby hyperfine transitions. This means that the zero-crossing of the (FM spectroscopy) error signal is shifted to a higher frequency. This also means that the zero-crossing depends on the optical probe power, so small power fluctuations from e.g.\ fiber couplings will cause fluctuations in the laser frequency. The laser frequency is locked slightly above the zero-crossing, for this reason, but the effect cannot be completely eliminated, for larger power fluctuations.

A third explanation is that etalon fringes created by reflective surfaces in the optical setup, cause slow spurious signals when the path lengths are changing due to e.g.\ small temperature fluctuations.

A fourth explanation is simply that the data acquisition system has drifting offsets for the voltage measurements.

Finally it should be noted that when we consider the full dynamic range of the magnetometer extra care must be taken, since carrier and higher order sidebands in probes 1--4 will probe the Doppler background, and nearby hyperfine transitions, to give small spurious signals. Similarly, light shifts from carrier and higher order sidebands might give systematic errors across the full dynamic range. This could impact the conclusions drawn from \cref{fig:07_EPI_recording_spatial}. To check that this is not the case, recordings where made with reduced gradient amplitudes of 75\%, 50\%, and 25\%, which all showed similar relative nonlinearities. Future work should aim at characterizing and compensating such infidelities, to give a reliable measurement across the entire dynamic range. Alternatively, modulation frequencies $\nu_i$ of about 10~GHz (and correspondingly lower $\nu_0$) could be employed: this would greatly reduce light shifts, and eliminate spurious signals from carrier and higher order sidebands, since the group of hyperfine transitions spans about 9~GHz \cite{Staerkind2023}.

All the above aside it should be noted that assigning an absolute accuracy to a prototype sensor, is a rather speculative task. The probe lock-in oscillator phase needs to be set upon each power up, and fiber couplings need regular readjustment for consistent performance. Ultimately, several highly stable plug and play devices should be manufactured and calibrated at a certified metrology lab, before a meaningful absolute accuracy can be claimed.

\subsection{Comparison with NMR probes}
The impressive line of research and innovation that started with \cite{DeZanche2008,Barmet2008,Barmet2009,Wilm2011,Vannesjo2015} uses NMR probes for k-space trajectory correction, and MRI magnet characterization. This has evolved into a mature solution now deployed in different branches of MRI research \cite{Versteeg2022,Boulant2023,Feizollah2023}.

This technology has the advantage of very high sensitivity and accuracy, but suffers from issues related to rf interference and short measurement pulses that are even shorter in strong gradient fields. Also NMR probes are electronically tuned for a specific field strength, and can therefore not be used for, e.g., both a 7~T scanner and a 3~T scanner.

Optical probes as described in this work, doesn't suffer from any of these problems, and could --- when further developed --- provide a much more convenient solution for k-space trajectory correction in MRI. The disadvantages of the optical approach is currently the lower sensitivity and accuracy, which needs to be improved.

\subsection{Sensitivity and bandwidth limits}
With the data acquisition system currently limiting the sensitivity of the magnetometer, it is clear that there is room for improvement. Future work should seek to approach and investigate the fundamental limits of this technology. Ultimately the quantum shot noise of the probing light will be an inevitable barrier. Modifications to the optical setup could perhaps bypass even this limit by using squeezed probe light \cite{Polzik1992}. After all, the task is to determine the center frequency of the optical transition, which has a natural linewidth of 5.2 MHz \cite{Young1994,Steck2023}. The 3.9~\textmu T resolution achieved in this work corresponds to an optical frequency resolution of 0.055~MHz, i.e.\ about 1\% of the natural linewidth.

In the measurements presented here we have used a sampling rate of 40 kHz, which is slow enough that we don't need to worry about the response time of the VCOs, and synchronization of recorded error signals and VCO voltages. We note that the upper bandwidth limit for this method will be the FM modulation frequency, which must be similar to the linewidth of the transition. This is equivalent to recognizing that measuring a magnetic-field change much faster than the 30~ns decay time \cite{Young1994} of the atomic transition is not possible.

\section{Conclusion}
We have presented a novel quantum sensor --- the \mbox{EXAAQ} magnetometer. The prototype is a fairly robust and compact device, despite being the very first demonstration of this technology. We have found the resolution of the sensor to be sub-ppm, and the accuracy to be about 5 ppm. We note that this accuracy is already much better than the best commercially available Hall probes \cite{Keller_Metrolab,3MH6-E_Senis}, which are limited to around 100 ppm. We have identified a number of simple limitations in the setup, leaving straightforward opportunities for future improvements.

This type of magnetometer can be configured to work at different field strengths and the number of probes can be chosen as necessary. It can readily be used for any high-field measurement application where low interference, high sensitivity and accuracy, and high bandwidth is of importance.

We have tested the sensor in a 7~T MRI scanner and found that it already works well enough to be used as a nice tool for probing the MRI scanner. We have only shown measurements of short sequences, to make details visible, but we emphasize that a key feature of the technology is that it can measure uninterrupted, e.g.\ during entire MRI sequences of many minutes. While the imperfections of the MRI scanner revealed in this work are no surprise, and could also have been uncovered using calibration sequences of the MRI scanner, or external NMR probes, it should be appreciated that they have been found using a completely novel approach.

The fact that vibrational eddy currents can clearly be resolved with the prototype, highlights the potential for this technology. An updated calibration of the ecc system would likely reduce the eddy currents, but it would only be a matter of time before this calibration again would be outdated. It is also possible that eddy current characteristics depend on the temperature of the gradient coil system, and hence change during several hours of operation. With a permanently installed optical magnetic-field-monitoring system such time-consuming calibrations could in the future be unnecessary, and optimal performance of the MRI scanner would always be ensured.

\section{Outlook}
In the continuation of this work we will work to improve the prototype towards an even more mature device, with less drift and high fidelity across the entire dynamic range. A number of steps can be taken to improve the current design. To deal with the temperature-dependent VCOs, different solutions could be employed: temperature stabilizing the VCOs; measuring the output frequencies instead of the control voltages; or replacing the VCOs with digital synthesizers. The last two options would also remove the problem of the noisy data acquisition system, as the measured/programmed frequency would be used directly.

While on the one hand working to improve the prototype, we will also start exploring k-space trajectory correction using the measured field data. A number of technical challenges must be solved in this regard: The sampling rate should ideally be increased to a couple of hundreds of kilohertz, accurate spatial localization of the probes should be realized, and measurements should be precisely synchronized with the MRI acquisition \cite{DeZanche2008,Barmet2008}.

For applications beyond MRI we note that reducing the probe size should be possible, since no particular attention has been paid to miniaturization, beyond what was necessary for installation in the MRI scanner bore. Expanding the dynamic range to several teslas could be realized using modulation frequencies $\nu_i$ in the microwave range. For applications where only a very narrow dynamic range is necessary, the probe feedback can be omitted --- equivalent to setting $\nu_i=0$ --- to increase the sensitivity and accuracy, and reducing complexity. It should be noted that measurements below 0.5~T could be challenging due to the high density of different transitions \cite{Staerkind2023}. Optical pumping on the D\textsubscript{1} line could be a solution. Also, $\pi$-transitions could cause spurious signals below 1~T, for imperfect circular probe polarization, or probes not perfectly aligned with the field direction.

The data sets and scripts for the analysis and calculations
underlying this work are openly available from \cite{Data}.

\section*{Acknowledgments}
We would like to thank Rikke H. L{\"u}tge and Axel Boisen for their contributions to the electronics development, J{\"u}rgen Appel for useful suggestions on the measurement strategy, and Michal Povazan, Kristin Engel, and Paul Sanders for assisting with the MRI system.
This project has received funding from the Danish Quantum Innovation Center (Qubiz)/Innovation Fund Denmark, the European Union’s Horizon 2020 research and innovation programme under Grant Agreements No. 820393 and No. 787520, and Villum Fonden under a Villum Investigator Grant, Grant No. 25880. The 7 T scanner was donated by the John and Birthe Meyer Foundation and The Danish Agency for Science, Technology and Innovation (Grant No. 0601-01370B).
\bibliography{Bibliography}

\begin{thebibliography}{55}%
\makeatletter
\providecommand \@ifxundefined [1]{%
 \@ifx{#1\undefined}
}%
\providecommand \@ifnum [1]{%
 \ifnum #1\expandafter \@firstoftwo
 \else \expandafter \@secondoftwo
 \fi
}%
\providecommand \@ifx [1]{%
 \ifx #1\expandafter \@firstoftwo
 \else \expandafter \@secondoftwo
 \fi
}%
\providecommand \natexlab [1]{#1}%
\providecommand \enquote  [1]{``#1''}%
\providecommand \bibnamefont  [1]{#1}%
\providecommand \bibfnamefont [1]{#1}%
\providecommand \citenamefont [1]{#1}%
\providecommand \href@noop [0]{\@secondoftwo}%
\providecommand \href [0]{\begingroup \@sanitize@url \@href}%
\providecommand \@href[1]{\@@startlink{#1}\@@href}%
\providecommand \@@href[1]{\endgroup#1\@@endlink}%
\providecommand \@sanitize@url [0]{\catcode `\\12\catcode `\$12\catcode `\&12\catcode `\#12\catcode `\^12\catcode `\_12\catcode `\%12\relax}%
\providecommand \@@startlink[1]{}%
\providecommand \@@endlink[0]{}%
\providecommand \url  [0]{\begingroup\@sanitize@url \@url }%
\providecommand \@url [1]{\endgroup\@href {#1}{\urlprefix }}%
\providecommand \urlprefix  [0]{URL }%
\providecommand \Eprint [0]{\href }%
\providecommand \doibase [0]{https://doi.org/}%
\providecommand \selectlanguage [0]{\@gobble}%
\providecommand \bibinfo  [0]{\@secondoftwo}%
\providecommand \bibfield  [0]{\@secondoftwo}%
\providecommand \translation [1]{[#1]}%
\providecommand \BibitemOpen [0]{}%
\providecommand \bibitemStop [0]{}%
\providecommand \bibitemNoStop [0]{.\EOS\space}%
\providecommand \EOS [0]{\spacefactor3000\relax}%
\providecommand \BibitemShut  [1]{\csname bibitem#1\endcsname}%
\let\auto@bib@innerbib\@empty
\bibitem [{\citenamefont {Battesti}\ \emph {et~al.}(2018)\citenamefont {Battesti}, \citenamefont {Beard}, \citenamefont {B{\"o}ser}, \citenamefont {Bruyant}, \citenamefont {Budker}, \citenamefont {Crooker}, \citenamefont {Daw}, \citenamefont {Flambaum}, \citenamefont {Inada}, \citenamefont {Irastorza} \emph {et~al.}}]{Battesti2018}%
  \BibitemOpen
  \bibfield  {author} {\bibinfo {author} {\bibfnamefont {R.}~\bibnamefont {Battesti}}, \bibinfo {author} {\bibfnamefont {J.}~\bibnamefont {Beard}}, \bibinfo {author} {\bibfnamefont {S.}~\bibnamefont {B{\"o}ser}}, \bibinfo {author} {\bibfnamefont {N.}~\bibnamefont {Bruyant}}, \bibinfo {author} {\bibfnamefont {D.}~\bibnamefont {Budker}}, \bibinfo {author} {\bibfnamefont {S.~A.}\ \bibnamefont {Crooker}}, \bibinfo {author} {\bibfnamefont {E.~J.}\ \bibnamefont {Daw}}, \bibinfo {author} {\bibfnamefont {V.~V.}\ \bibnamefont {Flambaum}}, \bibinfo {author} {\bibfnamefont {T.}~\bibnamefont {Inada}}, \bibinfo {author} {\bibfnamefont {I.~G.}\ \bibnamefont {Irastorza}}, \emph {et~al.},\ }\bibfield  {title} {\bibinfo {title} {\textit{High Magnetic Fields for Fundamental Physics}},\ }\href {https://doi.org/10.1016/j.physrep.2018.07.005} {\bibfield  {journal} {\bibinfo  {journal} {Physics Reports}\ }\textbf {\bibinfo {volume} {765-766}},\ \bibinfo {pages} {1} (\bibinfo {year} {2018})}\BibitemShut {NoStop}%
\bibitem [{\citenamefont {Roy}(2017)}]{Roy2017}%
  \BibitemOpen
  \bibfield  {author} {\bibinfo {author} {\bibfnamefont {S.}~\bibnamefont {Roy}},\ }\bibfield  {title} {\bibinfo {title} {\textit{Chapter 21 - Materials in a High Magnetic Field}},\ }in\ \href {https://doi.org/10.1016/B978-0-12-801300-7.00021-8} {\emph {\bibinfo {booktitle} {Materials Under Extreme Conditions}}},\ \bibinfo {editor} {edited by\ \bibinfo {editor} {\bibfnamefont {A.}~\bibnamefont {Tyagi}}\ and\ \bibinfo {editor} {\bibfnamefont {S.}~\bibnamefont {Banerjee}}}\ (\bibinfo  {publisher} {Elsevier},\ \bibinfo {year} {2017})\ pp.\ \bibinfo {pages} {755--789}\BibitemShut {NoStop}%
\bibitem [{\citenamefont {Caridad}\ \emph {et~al.}(2018)\citenamefont {Caridad}, \citenamefont {Power}, \citenamefont {Lotz}, \citenamefont {Shylau}, \citenamefont {Thomsen}, \citenamefont {Gammelgaard}, \citenamefont {Booth}, \citenamefont {Jauho},\ and\ \citenamefont {B{\o}ggild}}]{Caridad2018}%
  \BibitemOpen
  \bibfield  {author} {\bibinfo {author} {\bibfnamefont {J.~M.}\ \bibnamefont {Caridad}}, \bibinfo {author} {\bibfnamefont {S.~R.}\ \bibnamefont {Power}}, \bibinfo {author} {\bibfnamefont {M.~R.}\ \bibnamefont {Lotz}}, \bibinfo {author} {\bibfnamefont {A.~A.}\ \bibnamefont {Shylau}}, \bibinfo {author} {\bibfnamefont {J.~D.}\ \bibnamefont {Thomsen}}, \bibinfo {author} {\bibfnamefont {L.}~\bibnamefont {Gammelgaard}}, \bibinfo {author} {\bibfnamefont {T.~J.}\ \bibnamefont {Booth}}, \bibinfo {author} {\bibfnamefont {A.-P.}\ \bibnamefont {Jauho}},\ and\ \bibinfo {author} {\bibfnamefont {P.}~\bibnamefont {B{\o}ggild}},\ }\bibfield  {title} {\bibinfo {title} {\textit{Conductance quantization suppression in the quantum Hall regime}},\ }\href {https://doi.org/10.1038/s41467-018-03064-8} {\bibfield  {journal} {\bibinfo  {journal} {Nature Communications}\ }\textbf {\bibinfo {volume} {9}},\ \bibinfo {pages} {659} (\bibinfo {year} {2018})}\BibitemShut {NoStop}%
\bibitem [{\citenamefont {Marshall}\ \emph {et~al.}(1998)\citenamefont {Marshall}, \citenamefont {Hendrickson},\ and\ \citenamefont {Jackson}}]{Marshall1998}%
  \BibitemOpen
  \bibfield  {author} {\bibinfo {author} {\bibfnamefont {A.~G.}\ \bibnamefont {Marshall}}, \bibinfo {author} {\bibfnamefont {C.~L.}\ \bibnamefont {Hendrickson}},\ and\ \bibinfo {author} {\bibfnamefont {G.~S.}\ \bibnamefont {Jackson}},\ }\bibfield  {title} {\bibinfo {title} {\textit{Fourier transform ion cyclotron resonance mass spectrometry: A primer}},\ }\href {https://doi.org/10.1002/(SICI)1098-2787(1998)17:1<1::AID-MAS1>3.0.CO;2-K} {\bibfield  {journal} {\bibinfo  {journal} {Mass Spectrometry Reviews}\ }\textbf {\bibinfo {volume} {17}},\ \bibinfo {pages} {1} (\bibinfo {year} {1998})}\BibitemShut {NoStop}%
\bibitem [{\citenamefont {Gorshkov}\ \emph {et~al.}(1998)\citenamefont {Gorshkov}, \citenamefont {{Paša Tolić}}, \citenamefont {Udseth}, \citenamefont {Anderson}, \citenamefont {Huang}, \citenamefont {Bruce}, \citenamefont {Prior}, \citenamefont {Hofstadler}, \citenamefont {Tang}, \citenamefont {Chen} \emph {et~al.}}]{Gorshkov1998}%
  \BibitemOpen
  \bibfield  {author} {\bibinfo {author} {\bibfnamefont {M.~V.}\ \bibnamefont {Gorshkov}}, \bibinfo {author} {\bibfnamefont {L.}~\bibnamefont {{Paša Tolić}}}, \bibinfo {author} {\bibfnamefont {H.~R.}\ \bibnamefont {Udseth}}, \bibinfo {author} {\bibfnamefont {G.~A.}\ \bibnamefont {Anderson}}, \bibinfo {author} {\bibfnamefont {B.~M.}\ \bibnamefont {Huang}}, \bibinfo {author} {\bibfnamefont {J.~E.}\ \bibnamefont {Bruce}}, \bibinfo {author} {\bibfnamefont {D.~C.}\ \bibnamefont {Prior}}, \bibinfo {author} {\bibfnamefont {S.~A.}\ \bibnamefont {Hofstadler}}, \bibinfo {author} {\bibfnamefont {L.}~\bibnamefont {Tang}}, \bibinfo {author} {\bibfnamefont {L.-Z.}\ \bibnamefont {Chen}}, \emph {et~al.},\ }\bibfield  {title} {\bibinfo {title} {\textit{Electrospray Ionization–Fourier Transform Ion Cyclotron Resonance Mass Spectrometry at 11.5 Tesla: Instrument Design and Initial Results}},\ }\href {https://doi.org/10.1016/S1044-0305(98)00037-3} {\bibfield  {journal} {\bibinfo  {journal} {Journal of the American Society
  for Mass Spectrometry}\ }\textbf {\bibinfo {volume} {9}},\ \bibinfo {pages} {692} (\bibinfo {year} {1998})}\BibitemShut {NoStop}%
\bibitem [{\citenamefont {Bottura}\ \emph {et~al.}(2012)\citenamefont {Bottura}, \citenamefont {de~Rijk}, \citenamefont {Rossi},\ and\ \citenamefont {Todesco}}]{Bottura2012}%
  \BibitemOpen
  \bibfield  {author} {\bibinfo {author} {\bibfnamefont {L.}~\bibnamefont {Bottura}}, \bibinfo {author} {\bibfnamefont {G.}~\bibnamefont {de~Rijk}}, \bibinfo {author} {\bibfnamefont {L.}~\bibnamefont {Rossi}},\ and\ \bibinfo {author} {\bibfnamefont {E.}~\bibnamefont {Todesco}},\ }\bibfield  {title} {\bibinfo {title} {\textit{Advanced Accelerator Magnets for Upgrading the LHC}},\ }\href {https://doi.org/10.1109/TASC.2012.2186109} {\bibfield  {journal} {\bibinfo  {journal} {IEEE Transactions on Applied Superconductivity}\ }\textbf {\bibinfo {volume} {22}},\ \bibinfo {pages} {4002008} (\bibinfo {year} {2012})}\BibitemShut {NoStop}%
\bibitem [{\citenamefont {Bottura}\ \emph {et~al.}(2016)\citenamefont {Bottura}, \citenamefont {Gourlay}, \citenamefont {Yamamoto},\ and\ \citenamefont {Zlobin}}]{Bottura2016}%
  \BibitemOpen
  \bibfield  {author} {\bibinfo {author} {\bibfnamefont {L.}~\bibnamefont {Bottura}}, \bibinfo {author} {\bibfnamefont {S.~A.}\ \bibnamefont {Gourlay}}, \bibinfo {author} {\bibfnamefont {A.}~\bibnamefont {Yamamoto}},\ and\ \bibinfo {author} {\bibfnamefont {A.~V.}\ \bibnamefont {Zlobin}},\ }\bibfield  {title} {\bibinfo {title} {\textit{Superconducting Magnets for Particle Accelerators}},\ }\href {https://doi.org/10.1109/TNS.2015.2485159} {\bibfield  {journal} {\bibinfo  {journal} {IEEE Transactions on Nuclear Science}\ }\textbf {\bibinfo {volume} {63}},\ \bibinfo {pages} {751} (\bibinfo {year} {2016})}\BibitemShut {NoStop}%
\bibitem [{\citenamefont {Savary}\ \emph {et~al.}(2010)\citenamefont {Savary}, \citenamefont {Bonito-Oliva}, \citenamefont {Gallix}, \citenamefont {Knaster}, \citenamefont {Koizumi}, \citenamefont {Mitchell}, \citenamefont {Nakajima}, \citenamefont {Okuno},\ and\ \citenamefont {Sborchia}}]{Savary2010}%
  \BibitemOpen
  \bibfield  {author} {\bibinfo {author} {\bibfnamefont {F.}~\bibnamefont {Savary}}, \bibinfo {author} {\bibfnamefont {A.}~\bibnamefont {Bonito-Oliva}}, \bibinfo {author} {\bibfnamefont {R.}~\bibnamefont {Gallix}}, \bibinfo {author} {\bibfnamefont {J.}~\bibnamefont {Knaster}}, \bibinfo {author} {\bibfnamefont {N.}~\bibnamefont {Koizumi}}, \bibinfo {author} {\bibfnamefont {N.}~\bibnamefont {Mitchell}}, \bibinfo {author} {\bibfnamefont {H.}~\bibnamefont {Nakajima}}, \bibinfo {author} {\bibfnamefont {K.}~\bibnamefont {Okuno}},\ and\ \bibinfo {author} {\bibfnamefont {C.}~\bibnamefont {Sborchia}},\ }\bibfield  {title} {\bibinfo {title} {\textit{Status Report on the Toroidal Field Coils for the ITER Project}},\ }\href {https://doi.org/10.1109/TASC.2010.2040914} {\bibfield  {journal} {\bibinfo  {journal} {IEEE Transactions on Applied Superconductivity}\ }\textbf {\bibinfo {volume} {20}},\ \bibinfo {pages} {381} (\bibinfo {year} {2010})}\BibitemShut {NoStop}%
\bibitem [{\citenamefont {Zhai}\ \emph {et~al.}(2021)\citenamefont {Zhai}, \citenamefont {{van der Laan}}, \citenamefont {Connolly},\ and\ \citenamefont {Kessel}}]{Zhai2021}%
  \BibitemOpen
  \bibfield  {author} {\bibinfo {author} {\bibfnamefont {Y.}~\bibnamefont {Zhai}}, \bibinfo {author} {\bibfnamefont {D.}~\bibnamefont {{van der Laan}}}, \bibinfo {author} {\bibfnamefont {P.}~\bibnamefont {Connolly}},\ and\ \bibinfo {author} {\bibfnamefont {C.}~\bibnamefont {Kessel}},\ }\bibfield  {title} {\bibinfo {title} {\textit{Conceptual design of HTS magnets for fusion nuclear science facility}},\ }\href {https://doi.org/10.1016/j.fusengdes.2021.112611} {\bibfield  {journal} {\bibinfo  {journal} {Fusion Engineering and Design}\ }\textbf {\bibinfo {volume} {168}},\ \bibinfo {pages} {112611} (\bibinfo {year} {2021})}\BibitemShut {NoStop}%
\bibitem [{\citenamefont {Berry}\ and\ \citenamefont {Geim}(1997)}]{Berry1997}%
  \BibitemOpen
  \bibfield  {author} {\bibinfo {author} {\bibfnamefont {M.~V.}\ \bibnamefont {Berry}}\ and\ \bibinfo {author} {\bibfnamefont {A.~K.}\ \bibnamefont {Geim}},\ }\bibfield  {title} {\bibinfo {title} {\textit{Of flying frogs and levitrons}},\ }\href {https://doi.org/10.1088/0143-0807/18/4/012} {\bibfield  {journal} {\bibinfo  {journal} {European Journal of Physics}\ }\textbf {\bibinfo {volume} {18}},\ \bibinfo {pages} {307} (\bibinfo {year} {1997})}\BibitemShut {NoStop}%
\bibitem [{\citenamefont {Wang}\ \emph {et~al.}(2002)\citenamefont {Wang}, \citenamefont {Wang}, \citenamefont {Zeng}, \citenamefont {Huang}, \citenamefont {Luo}, \citenamefont {Xu}, \citenamefont {Tang}, \citenamefont {Lin}, \citenamefont {Zhang}, \citenamefont {Ren} \emph {et~al.}}]{Wang2002}%
  \BibitemOpen
  \bibfield  {author} {\bibinfo {author} {\bibfnamefont {J.}~\bibnamefont {Wang}}, \bibinfo {author} {\bibfnamefont {S.}~\bibnamefont {Wang}}, \bibinfo {author} {\bibfnamefont {Y.}~\bibnamefont {Zeng}}, \bibinfo {author} {\bibfnamefont {H.}~\bibnamefont {Huang}}, \bibinfo {author} {\bibfnamefont {F.}~\bibnamefont {Luo}}, \bibinfo {author} {\bibfnamefont {Z.}~\bibnamefont {Xu}}, \bibinfo {author} {\bibfnamefont {Q.}~\bibnamefont {Tang}}, \bibinfo {author} {\bibfnamefont {G.}~\bibnamefont {Lin}}, \bibinfo {author} {\bibfnamefont {C.}~\bibnamefont {Zhang}}, \bibinfo {author} {\bibfnamefont {Z.}~\bibnamefont {Ren}}, \emph {et~al.},\ }\bibfield  {title} {\bibinfo {title} {\textit{The first man-loading high temperature superconducting Maglev test vehicle in the world}},\ }\href {https://doi.org/10.1016/S0921-4534(02)01548-4} {\bibfield  {journal} {\bibinfo  {journal} {Physica C: Superconductivity}\ }\textbf {\bibinfo {volume} {378-381}},\ \bibinfo {pages} {809} (\bibinfo {year} {2002})}\BibitemShut {NoStop}%
\bibitem [{\citenamefont {Floegel-Delor}\ \emph {et~al.}(2019)\citenamefont {Floegel-Delor}, \citenamefont {Schirrmeister}, \citenamefont {Riedel}, \citenamefont {Koenig}, \citenamefont {Kantarbar}, \citenamefont {Liebmann},\ and\ \citenamefont {Werfel}}]{Floegel-Delor2019}%
  \BibitemOpen
  \bibfield  {author} {\bibinfo {author} {\bibfnamefont {U.}~\bibnamefont {Floegel-Delor}}, \bibinfo {author} {\bibfnamefont {P.}~\bibnamefont {Schirrmeister}}, \bibinfo {author} {\bibfnamefont {T.}~\bibnamefont {Riedel}}, \bibinfo {author} {\bibfnamefont {R.}~\bibnamefont {Koenig}}, \bibinfo {author} {\bibfnamefont {V.}~\bibnamefont {Kantarbar}}, \bibinfo {author} {\bibfnamefont {M.}~\bibnamefont {Liebmann}},\ and\ \bibinfo {author} {\bibfnamefont {F.~N.}\ \bibnamefont {Werfel}},\ }\bibfield  {title} {\bibinfo {title} {\textit{Mobile HTS Bulk Devices as Enabling Ton-Force Technology for Maglev Trains}},\ }\href {https://doi.org/10.1109/TASC.2019.2897216} {\bibfield  {journal} {\bibinfo  {journal} {IEEE Transactions on Applied Superconductivity}\ }\textbf {\bibinfo {volume} {29}},\ \bibinfo {pages} {1} (\bibinfo {year} {2019})}\BibitemShut {NoStop}%
\bibitem [{\citenamefont {Bothwell}\ and\ \citenamefont {Griffin}(2011)}]{Bothwell2011}%
  \BibitemOpen
  \bibfield  {author} {\bibinfo {author} {\bibfnamefont {J.~H.~F.}\ \bibnamefont {Bothwell}}\ and\ \bibinfo {author} {\bibfnamefont {J.~L.}\ \bibnamefont {Griffin}},\ }\bibfield  {title} {\bibinfo {title} {\textit{An introduction to biological nuclear magnetic resonance spectroscopy}},\ }\href {https://doi.org/10.1111/j.1469-185X.2010.00157.x} {\bibfield  {journal} {\bibinfo  {journal} {Biological Reviews}\ }\textbf {\bibinfo {volume} {86}},\ \bibinfo {pages} {493} (\bibinfo {year} {2011})}\BibitemShut {NoStop}%
\bibitem [{\citenamefont {Wikus}\ \emph {et~al.}(2022)\citenamefont {Wikus}, \citenamefont {Frantz}, \citenamefont {Kümmerle},\ and\ \citenamefont {Vonlanthen}}]{Wikus2022}%
  \BibitemOpen
  \bibfield  {author} {\bibinfo {author} {\bibfnamefont {P.}~\bibnamefont {Wikus}}, \bibinfo {author} {\bibfnamefont {W.}~\bibnamefont {Frantz}}, \bibinfo {author} {\bibfnamefont {R.}~\bibnamefont {Kümmerle}},\ and\ \bibinfo {author} {\bibfnamefont {P.}~\bibnamefont {Vonlanthen}},\ }\bibfield  {title} {\bibinfo {title} {\textit{Commercial gigahertz-class NMR magnets}},\ }\href {https://doi.org/10.1088/1361-6668/ac4951} {\bibfield  {journal} {\bibinfo  {journal} {Superconductor Science and Technology}\ }\textbf {\bibinfo {volume} {35}},\ \bibinfo {pages} {033001} (\bibinfo {year} {2022})}\BibitemShut {NoStop}%
\bibitem [{\citenamefont {Harisinghani}\ \emph {et~al.}(2019)\citenamefont {Harisinghani}, \citenamefont {O’Shea},\ and\ \citenamefont {Weissleder}}]{Harisinghani2019}%
  \BibitemOpen
  \bibfield  {author} {\bibinfo {author} {\bibfnamefont {M.~G.}\ \bibnamefont {Harisinghani}}, \bibinfo {author} {\bibfnamefont {A.}~\bibnamefont {O’Shea}},\ and\ \bibinfo {author} {\bibfnamefont {R.}~\bibnamefont {Weissleder}},\ }\bibfield  {title} {\bibinfo {title} {\textit{Advances in clinical MRI technology}},\ }\href {https://doi.org/10.1126/scitranslmed.aba2591} {\bibfield  {journal} {\bibinfo  {journal} {Science Translational Medicine}\ }\textbf {\bibinfo {volume} {11}},\ \bibinfo {pages} {eaba2591} (\bibinfo {year} {2019})}\BibitemShut {NoStop}%
\bibitem [{\citenamefont {Lvovsky}\ \emph {et~al.}(2013)\citenamefont {Lvovsky}, \citenamefont {Stautner},\ and\ \citenamefont {Zhang}}]{Lvovsky2013}%
  \BibitemOpen
  \bibfield  {author} {\bibinfo {author} {\bibfnamefont {Y.}~\bibnamefont {Lvovsky}}, \bibinfo {author} {\bibfnamefont {E.~W.}\ \bibnamefont {Stautner}},\ and\ \bibinfo {author} {\bibfnamefont {T.}~\bibnamefont {Zhang}},\ }\bibfield  {title} {\bibinfo {title} {\textit{Novel technologies and configurations of superconducting magnets for MRI}},\ }\href {https://doi.org/10.1088/0953-2048/26/9/093001} {\bibfield  {journal} {\bibinfo  {journal} {Superconductor Science and Technology}\ }\textbf {\bibinfo {volume} {26}},\ \bibinfo {pages} {093001} (\bibinfo {year} {2013})}\BibitemShut {NoStop}%
\bibitem [{\citenamefont {Keller}()}]{Keller_Metrolab}%
  \BibitemOpen
  \bibfield  {author} {\bibinfo {author} {\bibfnamefont {P.}~\bibnamefont {Keller}},\ }\href@noop {} {\bibinfo {title} {\textit{Technologies for Precision Magnetic Field Mapping}}},\ \bibinfo {howpublished} {\url{https://www.metrolab.com/resources/downloads/}},\ \bibinfo {note} {retrieved August 18, 2023}\BibitemShut {NoStop}%
\bibitem [{\citenamefont {Bottura}(2009)}]{Bottura2009}%
  \BibitemOpen
  \bibfield  {author} {\bibinfo {author} {\bibfnamefont {L.}~\bibnamefont {Bottura}},\ }\href@noop {} {\bibinfo {title} {\textit{Field Measurement Methods}}},\ \bibinfo {howpublished} {\url{https://cas.web.cern.ch/sites/default/files/lectures/bruges-2009/bottura-1.pdf}} (\bibinfo {year} {2009}),\ \bibinfo {note} {retrieved September 3, 2023}\BibitemShut {NoStop}%
\bibitem [{\citenamefont {Lenz}\ and\ \citenamefont {Edelstein}(2006)}]{Lenz2006}%
  \BibitemOpen
  \bibfield  {author} {\bibinfo {author} {\bibfnamefont {J.}~\bibnamefont {Lenz}}\ and\ \bibinfo {author} {\bibfnamefont {S.}~\bibnamefont {Edelstein}},\ }\bibfield  {title} {\bibinfo {title} {\textit{Magnetic sensors and their applications}},\ }\href {https://doi.org/10.1109/JSEN.2006.874493} {\bibfield  {journal} {\bibinfo  {journal} {IEEE Sensors Journal}\ }\textbf {\bibinfo {volume} {6}},\ \bibinfo {pages} {631} (\bibinfo {year} {2006})}\BibitemShut {NoStop}%
\bibitem [{\citenamefont {De~Zanche}\ \emph {et~al.}(2008)\citenamefont {De~Zanche}, \citenamefont {Barmet}, \citenamefont {Nordmeyer-Massner},\ and\ \citenamefont {Pruessmann}}]{DeZanche2008}%
  \BibitemOpen
  \bibfield  {author} {\bibinfo {author} {\bibfnamefont {N.}~\bibnamefont {De~Zanche}}, \bibinfo {author} {\bibfnamefont {C.}~\bibnamefont {Barmet}}, \bibinfo {author} {\bibfnamefont {J.~A.}\ \bibnamefont {Nordmeyer-Massner}},\ and\ \bibinfo {author} {\bibfnamefont {K.~P.}\ \bibnamefont {Pruessmann}},\ }\bibfield  {title} {\bibinfo {title} {\textit{NMR Probes for Measuring Magnetic Fields and Field Dynamics in MR Systems}},\ }\href {https://doi.org/10.1002/mrm.21624} {\bibfield  {journal} {\bibinfo  {journal} {Magnetic Resonance in Medicine}\ }\textbf {\bibinfo {volume} {60}},\ \bibinfo {pages} {176} (\bibinfo {year} {2008})}\BibitemShut {NoStop}%
\bibitem [{\citenamefont {Metrolab}({\natexlab{a}})}]{PT2026_Metrolab}%
  \BibitemOpen
  \bibfield  {author} {\bibinfo {author} {\bibnamefont {Metrolab}},\ }\href@noop {} {\bibinfo {title} {\textit{PT2026 Key Specifications}}},\ \bibinfo {howpublished} {\url{https://www.metrolab.com/wp-content/uploads/2021/06/PT2026-Key-specifications-v1.2.pdf}} ({\natexlab{a}}),\ \bibinfo {note} {retrieved September 3, 2023}\BibitemShut {NoStop}%
\bibitem [{\citenamefont {Nikiel}\ \emph {et~al.}(2014)\citenamefont {Nikiel}, \citenamefont {Bl{\"u}mler}, \citenamefont {Heil}, \citenamefont {Hehn}, \citenamefont {Karpuk}, \citenamefont {Maul}, \citenamefont {Otten}, \citenamefont {Schreiber},\ and\ \citenamefont {Terekhov}}]{Nikiel2014}%
  \BibitemOpen
  \bibfield  {author} {\bibinfo {author} {\bibfnamefont {A.}~\bibnamefont {Nikiel}}, \bibinfo {author} {\bibfnamefont {P.}~\bibnamefont {Bl{\"u}mler}}, \bibinfo {author} {\bibfnamefont {W.}~\bibnamefont {Heil}}, \bibinfo {author} {\bibfnamefont {M.}~\bibnamefont {Hehn}}, \bibinfo {author} {\bibfnamefont {S.}~\bibnamefont {Karpuk}}, \bibinfo {author} {\bibfnamefont {A.}~\bibnamefont {Maul}}, \bibinfo {author} {\bibfnamefont {E.}~\bibnamefont {Otten}}, \bibinfo {author} {\bibfnamefont {L.~M.}\ \bibnamefont {Schreiber}},\ and\ \bibinfo {author} {\bibfnamefont {M.}~\bibnamefont {Terekhov}},\ }\bibfield  {title} {\bibinfo {title} {\textit{Ultrasensitive 3He Magnetometer for Measurements of High Magnetic Fields}},\ }\href {https://doi.org/10.1140/epjd/e2014-50401-3} {\bibfield  {journal} {\bibinfo  {journal} {The European Physical Journal D}\ }\textbf {\bibinfo {volume} {68}},\ \bibinfo {pages} {330} (\bibinfo {year} {2014})}\BibitemShut {NoStop}%
\bibitem [{\citenamefont {Barmet}\ \emph {et~al.}(2010)\citenamefont {Barmet}, \citenamefont {Wilm}, \citenamefont {Pavan}, \citenamefont {Katsikatsos}, \citenamefont {Keupp}, \citenamefont {Mens},\ and\ \citenamefont {Pruessmann}}]{Barmet2010}%
  \BibitemOpen
  \bibfield  {author} {\bibinfo {author} {\bibfnamefont {C.}~\bibnamefont {Barmet}}, \bibinfo {author} {\bibfnamefont {B.~J.}\ \bibnamefont {Wilm}}, \bibinfo {author} {\bibfnamefont {M.}~\bibnamefont {Pavan}}, \bibinfo {author} {\bibfnamefont {G.}~\bibnamefont {Katsikatsos}}, \bibinfo {author} {\bibfnamefont {J.}~\bibnamefont {Keupp}}, \bibinfo {author} {\bibfnamefont {G.}~\bibnamefont {Mens}},\ and\ \bibinfo {author} {\bibfnamefont {K.}~\bibnamefont {Pruessmann}},\ }\bibfield  {title} {\bibinfo {title} {\textit{Concurrent Higher-Order Field Monitoring for Routine Head MRI: An Integrated Heteronuclear Setup}},\ }\href@noop {} {\bibfield  {journal} {\bibinfo  {journal} {Proc. Intl. Soc. Mag. Reson. Med. 18}\ ,\ \bibinfo {pages} {216}} (\bibinfo {year} {2010})}\BibitemShut {NoStop}%
\bibitem [{\citenamefont {Senis}()}]{3MH6-E_Senis}%
  \BibitemOpen
  \bibfield  {author} {\bibinfo {author} {\bibnamefont {Senis}},\ }\href@noop {} {\bibinfo {title} {\textit{3MH6-E TESLAMETER}}},\ \bibinfo {howpublished} {\url{https://www.senis.swiss/wp-content/uploads/2023/04/DS.200.3MH6-E-TESLAMETER_Rev2.2.pdf}},\ \bibinfo {note} {retrieved September 4, 2023}\BibitemShut {NoStop}%
\bibitem [{\citenamefont {Pedersen}\ \emph {et~al.}(2019)\citenamefont {Pedersen}, \citenamefont {Hanson}, \citenamefont {Xue},\ and\ \citenamefont {Hanson}}]{Pedersen2009}%
  \BibitemOpen
  \bibfield  {author} {\bibinfo {author} {\bibfnamefont {J.~O.}\ \bibnamefont {Pedersen}}, \bibinfo {author} {\bibfnamefont {C.~G.}\ \bibnamefont {Hanson}}, \bibinfo {author} {\bibfnamefont {R.}~\bibnamefont {Xue}},\ and\ \bibinfo {author} {\bibfnamefont {L.~G.}\ \bibnamefont {Hanson}},\ }\bibfield  {title} {\bibinfo {title} {\textit{Inductive Measurement and Encoding of k-space Trajectories in MR Raw Data}},\ }\href {https://doi.org/10.1007/s10334-019-00770-2} {\bibfield  {journal} {\bibinfo  {journal} {Magnetic Resonance Materials in Physics, Biology and Medicine}\ }\textbf {\bibinfo {volume} {32}},\ \bibinfo {pages} {655} (\bibinfo {year} {2019})}\BibitemShut {NoStop}%
\bibitem [{\citenamefont {Adaikkan}\ \emph {et~al.}(2022)\citenamefont {Adaikkan}, \citenamefont {Ma}, \citenamefont {Walach}, \citenamefont {Bochard}, \citenamefont {Bechtold}, \citenamefont {Vayakis}, \citenamefont {Walsh}, \citenamefont {Lino},\ and\ \citenamefont {Counsell}}]{Adaikkan2022}%
  \BibitemOpen
  \bibfield  {author} {\bibinfo {author} {\bibfnamefont {M.}~\bibnamefont {Adaikkan}}, \bibinfo {author} {\bibfnamefont {Y.}~\bibnamefont {Ma}}, \bibinfo {author} {\bibfnamefont {U.}~\bibnamefont {Walach}}, \bibinfo {author} {\bibfnamefont {M.}~\bibnamefont {Bochard}}, \bibinfo {author} {\bibfnamefont {F.}~\bibnamefont {Bechtold}}, \bibinfo {author} {\bibfnamefont {G.}~\bibnamefont {Vayakis}}, \bibinfo {author} {\bibfnamefont {M.}~\bibnamefont {Walsh}}, \bibinfo {author} {\bibfnamefont {M.~P.~C.}\ \bibnamefont {Lino}},\ and\ \bibinfo {author} {\bibfnamefont {G.}~\bibnamefont {Counsell}},\ }\bibfield  {title} {\bibinfo {title} {\textit{Producing High Quality LTCC Sensors for ITER In-Vessel Magnetic Diagnostics}},\ }\href {https://doi.org/10.1016/j.fusengdes.2022.113316} {\bibfield  {journal} {\bibinfo  {journal} {Fusion Engineering and Design}\ }\textbf {\bibinfo {volume} {184}},\ \bibinfo {pages} {113316} (\bibinfo {year} {2022})}\BibitemShut {NoStop}%
\bibitem [{\citenamefont {Metrolab}({\natexlab{b}})}]{FDI2056_Metrolab}%
  \BibitemOpen
  \bibfield  {author} {\bibinfo {author} {\bibnamefont {Metrolab}},\ }\href@noop {} {\bibinfo {title} {\textit{FDI2056 The First Off-The-Shelf Instrument to Quantify Magnetic Field Transients}}},\ \bibinfo {howpublished} {\url{https://www.metrolab.com/wp-content/uploads/2019/06/FDI2056_V4.pdf}} ({\natexlab{b}}),\ \bibinfo {note} {retrieved September 4, 2023}\BibitemShut {NoStop}%
\bibitem [{\citenamefont {Nakamura}\ \emph {et~al.}(2013)\citenamefont {Nakamura}, \citenamefont {Sawabe}, \citenamefont {Matsuda},\ and\ \citenamefont {Takeyama}}]{Nakamura2013}%
  \BibitemOpen
  \bibfield  {author} {\bibinfo {author} {\bibfnamefont {D.}~\bibnamefont {Nakamura}}, \bibinfo {author} {\bibfnamefont {H.}~\bibnamefont {Sawabe}}, \bibinfo {author} {\bibfnamefont {Y.~H.}\ \bibnamefont {Matsuda}},\ and\ \bibinfo {author} {\bibfnamefont {S.}~\bibnamefont {Takeyama}},\ }\bibfield  {title} {\bibinfo {title} {\textit{Precise Measurement of a Magnetic Field Generated by the Electromagnetic Flux Compression Technique}},\ }\href {https://doi.org/10.1063/1.4798543} {\bibfield  {journal} {\bibinfo  {journal} {Review of Scientific Instruments}\ }\textbf {\bibinfo {volume} {84}},\ \bibinfo {pages} {044702} (\bibinfo {year} {2013})}\BibitemShut {NoStop}%
\bibitem [{\citenamefont {Tsuji-lio}\ \emph {et~al.}(2001)\citenamefont {Tsuji-lio}, \citenamefont {Akiyama}, \citenamefont {Sato}, \citenamefont {Nozawa}, \citenamefont {Tsutsui}, \citenamefont {Shimada}, \citenamefont {Takahashi},\ and\ \citenamefont {Terai}}]{Tsuli-lio2001}%
  \BibitemOpen
  \bibfield  {author} {\bibinfo {author} {\bibfnamefont {S.}~\bibnamefont {Tsuji-lio}}, \bibinfo {author} {\bibfnamefont {T.}~\bibnamefont {Akiyama}}, \bibinfo {author} {\bibfnamefont {E.}~\bibnamefont {Sato}}, \bibinfo {author} {\bibfnamefont {T.}~\bibnamefont {Nozawa}}, \bibinfo {author} {\bibfnamefont {H.}~\bibnamefont {Tsutsui}}, \bibinfo {author} {\bibfnamefont {R.}~\bibnamefont {Shimada}}, \bibinfo {author} {\bibfnamefont {M.}~\bibnamefont {Takahashi}},\ and\ \bibinfo {author} {\bibfnamefont {K.}~\bibnamefont {Terai}},\ }\bibfield  {title} {\bibinfo {title} {\textit{Fiberoptic Heterodyne Magnetic Field Sensor for Long-Pulsed Fusion Devices}},\ }\href {https://doi.org/10.1063/1.1316751} {\bibfield  {journal} {\bibinfo  {journal} {Review of Scientific Instruments}\ }\textbf {\bibinfo {volume} {72}},\ \bibinfo {pages} {413} (\bibinfo {year} {2001})}\BibitemShut {NoStop}%
\bibitem [{\citenamefont {George}\ \emph {et~al.}(2017)\citenamefont {George}, \citenamefont {Bruyant}, \citenamefont {B{\'e}ard}, \citenamefont {Scotto}, \citenamefont {Arimondo}, \citenamefont {Battesti}, \citenamefont {Ciampini},\ and\ \citenamefont {Rizzo}}]{George2017}%
  \BibitemOpen
  \bibfield  {author} {\bibinfo {author} {\bibfnamefont {S.}~\bibnamefont {George}}, \bibinfo {author} {\bibfnamefont {N.}~\bibnamefont {Bruyant}}, \bibinfo {author} {\bibfnamefont {J.}~\bibnamefont {B{\'e}ard}}, \bibinfo {author} {\bibfnamefont {S.}~\bibnamefont {Scotto}}, \bibinfo {author} {\bibfnamefont {E.}~\bibnamefont {Arimondo}}, \bibinfo {author} {\bibfnamefont {R.}~\bibnamefont {Battesti}}, \bibinfo {author} {\bibfnamefont {D.}~\bibnamefont {Ciampini}},\ and\ \bibinfo {author} {\bibfnamefont {C.}~\bibnamefont {Rizzo}},\ }\bibfield  {title} {\bibinfo {title} {\textit{Pulsed High Magnetic Field Measurement with a Rubidium Vapor Sensor}},\ }\href {https://doi.org/10.1063/1.4993760} {\bibfield  {journal} {\bibinfo  {journal} {Rev. Sci. Instrum.}\ }\textbf {\bibinfo {volume} {88}},\ \bibinfo {pages} {073102} (\bibinfo {year} {2017})}\BibitemShut {NoStop}%
\bibitem [{\citenamefont {Ciampini}\ \emph {et~al.}(2017)\citenamefont {Ciampini}, \citenamefont {Battesti}, \citenamefont {Rizzo},\ and\ \citenamefont {Arimondo}}]{Ciampini2017}%
  \BibitemOpen
  \bibfield  {author} {\bibinfo {author} {\bibfnamefont {D.}~\bibnamefont {Ciampini}}, \bibinfo {author} {\bibfnamefont {R.}~\bibnamefont {Battesti}}, \bibinfo {author} {\bibfnamefont {C.}~\bibnamefont {Rizzo}},\ and\ \bibinfo {author} {\bibfnamefont {E.}~\bibnamefont {Arimondo}},\ }\bibfield  {title} {\bibinfo {title} {\textit{Optical Spectroscopy of a Microsized Rb Vapor Sample in Magnetic Fields up to 58 T}},\ }\href {https://doi.org/10.1103/PhysRevA.96.052504} {\bibfield  {journal} {\bibinfo  {journal} {Phys. Rev. A}\ }\textbf {\bibinfo {volume} {96}},\ \bibinfo {pages} {052504} (\bibinfo {year} {2017})}\BibitemShut {NoStop}%
\bibitem [{\citenamefont {Keaveney}\ \emph {et~al.}(2019)\citenamefont {Keaveney}, \citenamefont {Ponciano-Ojeda}, \citenamefont {Rieche}, \citenamefont {Raine}, \citenamefont {Hampshire},\ and\ \citenamefont {Hughes}}]{Keaveney2019}%
  \BibitemOpen
  \bibfield  {author} {\bibinfo {author} {\bibfnamefont {J.}~\bibnamefont {Keaveney}}, \bibinfo {author} {\bibfnamefont {F.~S.}\ \bibnamefont {Ponciano-Ojeda}}, \bibinfo {author} {\bibfnamefont {S.~M.}\ \bibnamefont {Rieche}}, \bibinfo {author} {\bibfnamefont {M.~J.}\ \bibnamefont {Raine}}, \bibinfo {author} {\bibfnamefont {D.~P.}\ \bibnamefont {Hampshire}},\ and\ \bibinfo {author} {\bibfnamefont {I.~G.}\ \bibnamefont {Hughes}},\ }\bibfield  {title} {\bibinfo {title} {\textit{Quantitative Optical Spectroscopy of $^{87}$Rb Vapour in the Voigt Geometry in DC Magnetic Fields up to 0.4 T}},\ }\href {https://doi.org/10.1088/1361-6455/ab0186} {\bibfield  {journal} {\bibinfo  {journal} {J. Phys. B: At. Mol. Opt. Phys}\ }\textbf {\bibinfo {volume} {52}},\ \bibinfo {pages} {055003} (\bibinfo {year} {2019})}\BibitemShut {NoStop}%
\bibitem [{\citenamefont {Klinger}\ \emph {et~al.}(2020)\citenamefont {Klinger}, \citenamefont {Azizbekyan}, \citenamefont {Sargsyan}, \citenamefont {Leroy}, \citenamefont {Sarkisyan},\ and\ \citenamefont {Papoyan}}]{Klinger2020}%
  \BibitemOpen
  \bibfield  {author} {\bibinfo {author} {\bibfnamefont {E.}~\bibnamefont {Klinger}}, \bibinfo {author} {\bibfnamefont {H.}~\bibnamefont {Azizbekyan}}, \bibinfo {author} {\bibfnamefont {A.}~\bibnamefont {Sargsyan}}, \bibinfo {author} {\bibfnamefont {C.}~\bibnamefont {Leroy}}, \bibinfo {author} {\bibfnamefont {D.}~\bibnamefont {Sarkisyan}},\ and\ \bibinfo {author} {\bibfnamefont {A.}~\bibnamefont {Papoyan}},\ }\bibfield  {title} {\bibinfo {title} {\textit{Proof of the Feasibility of a Nanocell-Based Wide-Range Optical Magnetometer}},\ }\href {https://doi.org/10.1364/AO.373949} {\bibfield  {journal} {\bibinfo  {journal} {Appl. Opt.}\ }\textbf {\bibinfo {volume} {59}},\ \bibinfo {pages} {2231} (\bibinfo {year} {2020})}\BibitemShut {NoStop}%
\bibitem [{\citenamefont {St{\ae}rkind}\ \emph {et~al.}(2023)\citenamefont {St{\ae}rkind}, \citenamefont {Jensen}, \citenamefont {M{\"u}ller}, \citenamefont {Boer}, \citenamefont {Petersen},\ and\ \citenamefont {Polzik}}]{Staerkind2023}%
  \BibitemOpen
  \bibfield  {author} {\bibinfo {author} {\bibfnamefont {H.}~\bibnamefont {St{\ae}rkind}}, \bibinfo {author} {\bibfnamefont {K.}~\bibnamefont {Jensen}}, \bibinfo {author} {\bibfnamefont {J.~H.}\ \bibnamefont {M{\"u}ller}}, \bibinfo {author} {\bibfnamefont {V.~O.}\ \bibnamefont {Boer}}, \bibinfo {author} {\bibfnamefont {E.~T.}\ \bibnamefont {Petersen}},\ and\ \bibinfo {author} {\bibfnamefont {E.~S.}\ \bibnamefont {Polzik}},\ }\bibfield  {title} {\bibinfo {title} {\textit{Precision Measurement of the Excited State Land\'e g-factor and Diamagnetic Shift of the Cesium ${\mathrm{D}}_{2}$ Line}},\ }\href {https://doi.org/10.1103/PhysRevX.13.021036} {\bibfield  {journal} {\bibinfo  {journal} {Phys. Rev. X}\ }\textbf {\bibinfo {volume} {13}},\ \bibinfo {pages} {021036} (\bibinfo {year} {2023})}\BibitemShut {NoStop}%
\bibitem [{\citenamefont {Br{\"a}uer}\ and\ \citenamefont {Budde}(2012)}]{Brauer2012}%
  \BibitemOpen
  \bibfield  {author} {\bibinfo {author} {\bibfnamefont {M.}~\bibnamefont {Br{\"a}uer}}\ and\ \bibinfo {author} {\bibfnamefont {M.}~\bibnamefont {Budde}},\ }\href {https://patents.google.com/patent/DE102012202237B4/en} {\bibinfo {title} {\textit{Vorrichtung und Verfahren zur Magnetfeldmessung und -regelung - DE102012202237B4}}} (\bibinfo {year} {2012})\BibitemShut {NoStop}%
\bibitem [{\citenamefont {Barmet}\ \emph {et~al.}(2008)\citenamefont {Barmet}, \citenamefont {Zanche},\ and\ \citenamefont {Pruessmann}}]{Barmet2008}%
  \BibitemOpen
  \bibfield  {author} {\bibinfo {author} {\bibfnamefont {C.}~\bibnamefont {Barmet}}, \bibinfo {author} {\bibfnamefont {N.~D.}\ \bibnamefont {Zanche}},\ and\ \bibinfo {author} {\bibfnamefont {K.~P.}\ \bibnamefont {Pruessmann}},\ }\bibfield  {title} {\bibinfo {title} {\textit{Spatiotemporal Magnetic Field Monitoring for MR}},\ }\href {https://doi.org/10.1002/mrm.21603} {\bibfield  {journal} {\bibinfo  {journal} {Magnetic Resonance in Medicine}\ }\textbf {\bibinfo {volume} {60}},\ \bibinfo {pages} {187} (\bibinfo {year} {2008})}\BibitemShut {NoStop}%
\bibitem [{\citenamefont {Strait}\ \emph {et~al.}(2008)\citenamefont {Strait}, \citenamefont {Fredrickson}, \citenamefont {Moret},\ and\ \citenamefont {Takechi}}]{Strait2008}%
  \BibitemOpen
  \bibfield  {author} {\bibinfo {author} {\bibfnamefont {E.~J.}\ \bibnamefont {Strait}}, \bibinfo {author} {\bibfnamefont {E.~D.}\ \bibnamefont {Fredrickson}}, \bibinfo {author} {\bibfnamefont {J.-M.}\ \bibnamefont {Moret}},\ and\ \bibinfo {author} {\bibfnamefont {M.}~\bibnamefont {Takechi}},\ }\bibfield  {title} {\bibinfo {title} {\textit{Chapter 2: Magnetic Diagnostics}},\ }\href {https://doi.org/10.13182/FST08-A1674} {\bibfield  {journal} {\bibinfo  {journal} {Fusion Science and Technology}\ }\textbf {\bibinfo {volume} {53}},\ \bibinfo {pages} {304} (\bibinfo {year} {2008})}\BibitemShut {NoStop}%
\bibitem [{\citenamefont {Quercia}\ \emph {et~al.}(2022)\citenamefont {Quercia}, \citenamefont {Pironti}, \citenamefont {Bolshakova}, \citenamefont {Holyaka}, \citenamefont {Duran}, \citenamefont {Murari},\ and\ \citenamefont {Contributors}}]{Quercia2022}%
  \BibitemOpen
  \bibfield  {author} {\bibinfo {author} {\bibfnamefont {A.}~\bibnamefont {Quercia}}, \bibinfo {author} {\bibfnamefont {A.}~\bibnamefont {Pironti}}, \bibinfo {author} {\bibfnamefont {I.}~\bibnamefont {Bolshakova}}, \bibinfo {author} {\bibfnamefont {R.}~\bibnamefont {Holyaka}}, \bibinfo {author} {\bibfnamefont {I.}~\bibnamefont {Duran}}, \bibinfo {author} {\bibfnamefont {A.}~\bibnamefont {Murari}},\ and\ \bibinfo {author} {\bibfnamefont {J.}~\bibnamefont {Contributors}},\ }\bibfield  {title} {\bibinfo {title} {\textit{Long Term Operation of the Radiation-Hard Hall Probes System and the Path Toward a High Performance Hybrid Magnetic Field Sensor}},\ }\href {https://doi.org/10.1088/1741-4326/ac8aad} {\bibfield  {journal} {\bibinfo  {journal} {Nuclear Fusion}\ }\textbf {\bibinfo {volume} {62}},\ \bibinfo {pages} {106032} (\bibinfo {year} {2022})}\BibitemShut {NoStop}%
\bibitem [{\citenamefont {Marchevsky}(2021)}]{Marchevsky2021}%
  \BibitemOpen
  \bibfield  {author} {\bibinfo {author} {\bibfnamefont {M.}~\bibnamefont {Marchevsky}},\ }\bibfield  {title} {\bibinfo {title} {\textit{Quench Detection and Protection for High-Temperature Superconductor Accelerator Magnets}},\ }\href {https://doi.org/10.3390/instruments5030027} {\bibfield  {journal} {\bibinfo  {journal} {Instruments}\ }\textbf {\bibinfo {volume} {5}},\ \bibinfo {pages} {27} (\bibinfo {year} {2021})}\BibitemShut {NoStop}%
\bibitem [{\citenamefont {Schmidt}\ \emph {et~al.}(1994)\citenamefont {Schmidt}, \citenamefont {Knaak}, \citenamefont {Wynands},\ and\ \citenamefont {Meschede}}]{Schmidt1994}%
  \BibitemOpen
  \bibfield  {author} {\bibinfo {author} {\bibfnamefont {O.}~\bibnamefont {Schmidt}}, \bibinfo {author} {\bibfnamefont {K.-M.}\ \bibnamefont {Knaak}}, \bibinfo {author} {\bibfnamefont {R.}~\bibnamefont {Wynands}},\ and\ \bibinfo {author} {\bibfnamefont {D.}~\bibnamefont {Meschede}},\ }\bibfield  {title} {\bibinfo {title} {\textit{Cesium Saturation Spectroscopy Revisited: How to Reverse Peaks and Observe Narrow Resonances}},\ }\href {https://doi.org/10.1007/BF01081167} {\bibfield  {journal} {\bibinfo  {journal} {Applied Physics B}\ }\textbf {\bibinfo {volume} {59}},\ \bibinfo {pages} {167} (\bibinfo {year} {1994})}\BibitemShut {NoStop}%
\bibitem [{\citenamefont {Bjorklund}\ \emph {et~al.}(1983)\citenamefont {Bjorklund}, \citenamefont {Levenson}, \citenamefont {Lenth},\ and\ \citenamefont {Ortiz}}]{Bjorklund1983}%
  \BibitemOpen
  \bibfield  {author} {\bibinfo {author} {\bibfnamefont {G.~C.}\ \bibnamefont {Bjorklund}}, \bibinfo {author} {\bibfnamefont {M.~D.}\ \bibnamefont {Levenson}}, \bibinfo {author} {\bibfnamefont {W.}~\bibnamefont {Lenth}},\ and\ \bibinfo {author} {\bibfnamefont {C.}~\bibnamefont {Ortiz}},\ }\bibfield  {title} {\bibinfo {title} {\textit{Frequency Modulation (FM) Spectroscopy}},\ }\href {https://doi.org/10.1007/BF00688820} {\bibfield  {journal} {\bibinfo  {journal} {Applied Physics B}\ }\textbf {\bibinfo {volume} {32}},\ \bibinfo {pages} {145} (\bibinfo {year} {1983})}\BibitemShut {NoStop}%
\bibitem [{\citenamefont {Young}\ \emph {et~al.}(1994)\citenamefont {Young}, \citenamefont {Hill}, \citenamefont {Sibener}, \citenamefont {Price}, \citenamefont {Tanner}, \citenamefont {Wieman},\ and\ \citenamefont {Leone}}]{Young1994}%
  \BibitemOpen
  \bibfield  {author} {\bibinfo {author} {\bibfnamefont {L.}~\bibnamefont {Young}}, \bibinfo {author} {\bibfnamefont {W.~T.}\ \bibnamefont {Hill}}, \bibinfo {author} {\bibfnamefont {S.~J.}\ \bibnamefont {Sibener}}, \bibinfo {author} {\bibfnamefont {S.~D.}\ \bibnamefont {Price}}, \bibinfo {author} {\bibfnamefont {C.~E.}\ \bibnamefont {Tanner}}, \bibinfo {author} {\bibfnamefont {C.~E.}\ \bibnamefont {Wieman}},\ and\ \bibinfo {author} {\bibfnamefont {S.~R.}\ \bibnamefont {Leone}},\ }\bibfield  {title} {\bibinfo {title} {\textit{Precision Lifetime Measurements of Cs 6p $^{2}$${\mathit{P}}_{1/2}$ and 6p $^{2}$${\mathit{P}}_{3/2}$ Levels by Single-Photon Counting}},\ }\href {https://doi.org/10.1103/PhysRevA.50.2174} {\bibfield  {journal} {\bibinfo  {journal} {Phys. Rev. A}\ }\textbf {\bibinfo {volume} {50}},\ \bibinfo {pages} {2174} (\bibinfo {year} {1994})}\BibitemShut {NoStop}%
\bibitem [{\citenamefont {Steck}(2023)}]{Steck2023}%
  \BibitemOpen
  \bibfield  {author} {\bibinfo {author} {\bibfnamefont {D.}~\bibnamefont {Steck}},\ }\href@noop {} {\bibinfo {title} {\textit{Cesium D Line Data}}},\ \bibinfo {howpublished} {\url{https://steck.us/alkalidata/cesiumnumbers.pdf}} (\bibinfo {year} {2023}),\ \bibinfo {note} {retrieved September 5, 2023}\BibitemShut {NoStop}%
\bibitem [{\citenamefont {Bernstein}\ \emph {et~al.}(2004)\citenamefont {Bernstein}, \citenamefont {King},\ and\ \citenamefont {Zhou}}]{Bernstein2004}%
  \BibitemOpen
  \bibfield  {author} {\bibinfo {author} {\bibfnamefont {M.~A.}\ \bibnamefont {Bernstein}}, \bibinfo {author} {\bibfnamefont {K.~F.}\ \bibnamefont {King}},\ and\ \bibinfo {author} {\bibfnamefont {X.~J.}\ \bibnamefont {Zhou}},\ }\href@noop {} {\emph {\bibinfo {title} {\textit{Handbook of MRI Pulse Sequences}}}}\ (\bibinfo  {publisher} {Elsevier Academic Press},\ \bibinfo {year} {2004})\BibitemShut {NoStop}%
\bibitem [{\citenamefont {Tiesinga}\ \emph {et~al.}(2021)\citenamefont {Tiesinga}, \citenamefont {Mohr}, \citenamefont {Newell},\ and\ \citenamefont {Taylor}}]{Tiesinga2021}%
  \BibitemOpen
  \bibfield  {author} {\bibinfo {author} {\bibfnamefont {E.}~\bibnamefont {Tiesinga}}, \bibinfo {author} {\bibfnamefont {P.~J.}\ \bibnamefont {Mohr}}, \bibinfo {author} {\bibfnamefont {D.~B.}\ \bibnamefont {Newell}},\ and\ \bibinfo {author} {\bibfnamefont {B.~N.}\ \bibnamefont {Taylor}},\ }\bibfield  {title} {\bibinfo {title} {\textit{CODATA Recommended Values of the Fundamental Physical Constants: 2018}},\ }\href {https://doi.org/10.1103/RevModPhys.93.025010} {\bibfield  {journal} {\bibinfo  {journal} {Rev. Mod. Phys.}\ }\textbf {\bibinfo {volume} {93}},\ \bibinfo {pages} {025010} (\bibinfo {year} {2021})}\BibitemShut {NoStop}%
\bibitem [{\citenamefont {Markl}\ \emph {et~al.}(2003)\citenamefont {Markl}, \citenamefont {Bammer}, \citenamefont {Alley}, \citenamefont {Elkins}, \citenamefont {Draney}, \citenamefont {Barnett}, \citenamefont {Moseley}, \citenamefont {Glover},\ and\ \citenamefont {Pelc}}]{Markl2003}%
  \BibitemOpen
  \bibfield  {author} {\bibinfo {author} {\bibfnamefont {M.}~\bibnamefont {Markl}}, \bibinfo {author} {\bibfnamefont {R.}~\bibnamefont {Bammer}}, \bibinfo {author} {\bibfnamefont {M.}~\bibnamefont {Alley}}, \bibinfo {author} {\bibfnamefont {C.}~\bibnamefont {Elkins}}, \bibinfo {author} {\bibfnamefont {M.}~\bibnamefont {Draney}}, \bibinfo {author} {\bibfnamefont {A.}~\bibnamefont {Barnett}}, \bibinfo {author} {\bibfnamefont {M.}~\bibnamefont {Moseley}}, \bibinfo {author} {\bibfnamefont {G.}~\bibnamefont {Glover}},\ and\ \bibinfo {author} {\bibfnamefont {N.}~\bibnamefont {Pelc}},\ }\bibfield  {title} {\bibinfo {title} {\textit{Generalized Reconstruction of Phase Contrast MRI: Analysis and Correction of the Effect of Gradient Field Distortions}},\ }\href {https://doi.org/10.1002/mrm.10582} {\bibfield  {journal} {\bibinfo  {journal} {Magnetic Resonance in Medicine}\ }\textbf {\bibinfo {volume} {50}},\ \bibinfo {pages} {791} (\bibinfo {year} {2003})}\BibitemShut {NoStop}%
\bibitem [{\citenamefont {Winkler}\ \emph {et~al.}(2018)\citenamefont {Winkler}, \citenamefont {Schmitt}, \citenamefont {Landes}, \citenamefont {{de Bever}}, \citenamefont {Wade}, \citenamefont {Alejski},\ and\ \citenamefont {Rutt}}]{Winkler2018}%
  \BibitemOpen
  \bibfield  {author} {\bibinfo {author} {\bibfnamefont {S.~A.}\ \bibnamefont {Winkler}}, \bibinfo {author} {\bibfnamefont {F.}~\bibnamefont {Schmitt}}, \bibinfo {author} {\bibfnamefont {H.}~\bibnamefont {Landes}}, \bibinfo {author} {\bibfnamefont {J.}~\bibnamefont {{de Bever}}}, \bibinfo {author} {\bibfnamefont {T.}~\bibnamefont {Wade}}, \bibinfo {author} {\bibfnamefont {A.}~\bibnamefont {Alejski}},\ and\ \bibinfo {author} {\bibfnamefont {B.~K.}\ \bibnamefont {Rutt}},\ }\bibfield  {title} {\bibinfo {title} {\textit{Gradient and Shim Technologies for Ultra High Field MRI}},\ }\href {https://doi.org/10.1016/j.neuroimage.2016.11.033} {\bibfield  {journal} {\bibinfo  {journal} {NeuroImage}\ }\textbf {\bibinfo {volume} {168}},\ \bibinfo {pages} {59} (\bibinfo {year} {2018})}\BibitemShut {NoStop}%
\bibitem [{\citenamefont {Barmet}\ \emph {et~al.}(2009)\citenamefont {Barmet}, \citenamefont {De~Zanche}, \citenamefont {Wilm},\ and\ \citenamefont {Pruessmann}}]{Barmet2009}%
  \BibitemOpen
  \bibfield  {author} {\bibinfo {author} {\bibfnamefont {C.}~\bibnamefont {Barmet}}, \bibinfo {author} {\bibfnamefont {N.}~\bibnamefont {De~Zanche}}, \bibinfo {author} {\bibfnamefont {B.~J.}\ \bibnamefont {Wilm}},\ and\ \bibinfo {author} {\bibfnamefont {K.~P.}\ \bibnamefont {Pruessmann}},\ }\bibfield  {title} {\bibinfo {title} {\textit{A Transmit/Receive System for Magnetic Field Monitoring of In Vivo MRI}},\ }\href {https://doi.org/10.1002/mrm.21996} {\bibfield  {journal} {\bibinfo  {journal} {Magnetic Resonance in Medicine}\ }\textbf {\bibinfo {volume} {62}},\ \bibinfo {pages} {269} (\bibinfo {year} {2009})}\BibitemShut {NoStop}%
\bibitem [{\citenamefont {Wilm}\ \emph {et~al.}(2011)\citenamefont {Wilm}, \citenamefont {Barmet}, \citenamefont {Pavan},\ and\ \citenamefont {Pruessmann}}]{Wilm2011}%
  \BibitemOpen
  \bibfield  {author} {\bibinfo {author} {\bibfnamefont {B.~J.}\ \bibnamefont {Wilm}}, \bibinfo {author} {\bibfnamefont {C.}~\bibnamefont {Barmet}}, \bibinfo {author} {\bibfnamefont {M.}~\bibnamefont {Pavan}},\ and\ \bibinfo {author} {\bibfnamefont {K.~P.}\ \bibnamefont {Pruessmann}},\ }\bibfield  {title} {\bibinfo {title} {\textit{Higher Order Reconstruction for MRI in the Presence of Spatiotemporal Field Perturbations}},\ }\href {https://doi.org/10.1002/mrm.22767} {\bibfield  {journal} {\bibinfo  {journal} {Magnetic Resonance in Medicine}\ }\textbf {\bibinfo {volume} {65}},\ \bibinfo {pages} {1690} (\bibinfo {year} {2011})}\BibitemShut {NoStop}%
\bibitem [{\citenamefont {Vannesjo}\ \emph {et~al.}(2015)\citenamefont {Vannesjo}, \citenamefont {Wilm}, \citenamefont {Duerst}, \citenamefont {Gross}, \citenamefont {Brunner}, \citenamefont {Dietrich}, \citenamefont {Schmid}, \citenamefont {Barmet},\ and\ \citenamefont {Pruessmann}}]{Vannesjo2015}%
  \BibitemOpen
  \bibfield  {author} {\bibinfo {author} {\bibfnamefont {S.~J.}\ \bibnamefont {Vannesjo}}, \bibinfo {author} {\bibfnamefont {B.~J.}\ \bibnamefont {Wilm}}, \bibinfo {author} {\bibfnamefont {Y.}~\bibnamefont {Duerst}}, \bibinfo {author} {\bibfnamefont {S.}~\bibnamefont {Gross}}, \bibinfo {author} {\bibfnamefont {D.~O.}\ \bibnamefont {Brunner}}, \bibinfo {author} {\bibfnamefont {B.~E.}\ \bibnamefont {Dietrich}}, \bibinfo {author} {\bibfnamefont {T.}~\bibnamefont {Schmid}}, \bibinfo {author} {\bibfnamefont {C.}~\bibnamefont {Barmet}},\ and\ \bibinfo {author} {\bibfnamefont {K.~P.}\ \bibnamefont {Pruessmann}},\ }\bibfield  {title} {\bibinfo {title} {\textit{Retrospective Correction of Physiological Field Fluctuations in High-field Brain MRI Using Concurrent Field Monitoring}},\ }\href {https://doi.org/10.1002/mrm.25303} {\bibfield  {journal} {\bibinfo  {journal} {Magnetic Resonance in Medicine}\ }\textbf {\bibinfo {volume} {73}},\ \bibinfo {pages} {1833} (\bibinfo {year} {2015})}\BibitemShut {NoStop}%
\bibitem [{\citenamefont {Versteeg}\ \emph {et~al.}(2022)\citenamefont {Versteeg}, \citenamefont {Klomp},\ and\ \citenamefont {Siero}}]{Versteeg2022}%
  \BibitemOpen
  \bibfield  {author} {\bibinfo {author} {\bibfnamefont {E.}~\bibnamefont {Versteeg}}, \bibinfo {author} {\bibfnamefont {D.~W.~J.}\ \bibnamefont {Klomp}},\ and\ \bibinfo {author} {\bibfnamefont {J.~C.~W.}\ \bibnamefont {Siero}},\ }\bibfield  {title} {\bibinfo {title} {\textit{Accelerating Brain Imaging Using a Silent Spatial Encoding Axis}},\ }\href {https://doi.org/10.1002/mrm.29350} {\bibfield  {journal} {\bibinfo  {journal} {Magnetic Resonance in Medicine}\ }\textbf {\bibinfo {volume} {88}},\ \bibinfo {pages} {1785} (\bibinfo {year} {2022})}\BibitemShut {NoStop}%
\bibitem [{\citenamefont {Boulant}\ \emph {et~al.}(2023)\citenamefont {Boulant}, \citenamefont {Quettier}, \citenamefont {Aubert}, \citenamefont {Amadon}, \citenamefont {Belorgey}, \citenamefont {Berriaud}, \citenamefont {Bonnelye}, \citenamefont {Bredy}, \citenamefont {Chazel}, \citenamefont {Dilasser} \emph {et~al.}}]{Boulant2023}%
  \BibitemOpen
  \bibfield  {author} {\bibinfo {author} {\bibfnamefont {N.}~\bibnamefont {Boulant}}, \bibinfo {author} {\bibfnamefont {L.}~\bibnamefont {Quettier}}, \bibinfo {author} {\bibfnamefont {G.}~\bibnamefont {Aubert}}, \bibinfo {author} {\bibfnamefont {A.}~\bibnamefont {Amadon}}, \bibinfo {author} {\bibfnamefont {J.}~\bibnamefont {Belorgey}}, \bibinfo {author} {\bibfnamefont {C.}~\bibnamefont {Berriaud}}, \bibinfo {author} {\bibfnamefont {C.}~\bibnamefont {Bonnelye}}, \bibinfo {author} {\bibfnamefont {P.}~\bibnamefont {Bredy}}, \bibinfo {author} {\bibfnamefont {E.}~\bibnamefont {Chazel}}, \bibinfo {author} {\bibfnamefont {G.}~\bibnamefont {Dilasser}}, \emph {et~al.},\ }\bibfield  {title} {\bibinfo {title} {\textit{Commissioning of the Iseult CEA 11.7~T Whole-body MRI: Current Status, Gradient-magnet Interaction Tests and First Imaging Experience}},\ }\href {https://doi.org/10.1007/s10334-023-01063-5} {\bibfield  {journal} {\bibinfo  {journal} {Magnetic Resonance Materials in Physics, Biology and Medicine}\ }\textbf
  {\bibinfo {volume} {36}},\ \bibinfo {pages} {175} (\bibinfo {year} {2023})}\BibitemShut {NoStop}%
\bibitem [{\citenamefont {Feizollah}\ and\ \citenamefont {Tardif}(2023)}]{Feizollah2023}%
  \BibitemOpen
  \bibfield  {author} {\bibinfo {author} {\bibfnamefont {S.}~\bibnamefont {Feizollah}}\ and\ \bibinfo {author} {\bibfnamefont {C.~L.}\ \bibnamefont {Tardif}},\ }\bibfield  {title} {\bibinfo {title} {\textit{High-resolution Diffusion-weighted Imaging at 7 Tesla: Single-shot Readout Trajectories and Their Impact on Signal-to-noise Ratio, Spatial Resolution and Accuracy}},\ }\href {https://doi.org/10.1016/j.neuroimage.2023.120159} {\bibfield  {journal} {\bibinfo  {journal} {NeuroImage}\ }\textbf {\bibinfo {volume} {274}},\ \bibinfo {pages} {120159} (\bibinfo {year} {2023})}\BibitemShut {NoStop}%
\bibitem [{\citenamefont {Polzik}\ \emph {et~al.}(1992)\citenamefont {Polzik}, \citenamefont {Carri},\ and\ \citenamefont {Kimble}}]{Polzik1992}%
  \BibitemOpen
  \bibfield  {author} {\bibinfo {author} {\bibfnamefont {E.~S.}\ \bibnamefont {Polzik}}, \bibinfo {author} {\bibfnamefont {J.}~\bibnamefont {Carri}},\ and\ \bibinfo {author} {\bibfnamefont {H.~J.}\ \bibnamefont {Kimble}},\ }\bibfield  {title} {\bibinfo {title} {\textit{Spectroscopy with Squeezed Light}},\ }\href {https://doi.org/10.1103/PhysRevLett.68.3020} {\bibfield  {journal} {\bibinfo  {journal} {Phys. Rev. Lett.}\ }\textbf {\bibinfo {volume} {68}},\ \bibinfo {pages} {3020} (\bibinfo {year} {1992})}\BibitemShut {NoStop}%
\bibitem [{\citenamefont {St{\ae}rkind}(2023)}]{Data}%
  \BibitemOpen
  \bibfield  {author} {\bibinfo {author} {\bibfnamefont {H.}~\bibnamefont {St{\ae}rkind}},\ }\href@noop {} {\bibinfo {title} {\textit{Data underlying: High-Field Optical Cesium Magnetometer for Magnetic Resonance Imaging}}},\ \bibinfo {howpublished} {\url{https://doi.org/10.5281/zenodo.8366717}} (\bibinfo {year} {2023})\BibitemShut {NoStop}%
\end{thebibliography}%

\end{document}